# A Review of Vibration-Based Damage Detection in Civil Structures: From Traditional Methods to Machine Learning and Deep Learning Applications

Onur Avci[1], Osama Abdeljaber[2], Serkan Kiranyaz[3], Mohammed Hussein[4], Moncef Gabbouj[5], Daniel J. Inman[6]

[1] Guest Lecturer, School of Civil Engineering, University of Leeds, Leeds, United Kingdom.
Email: o.avci@leeds.ac.uk
[2] Postdoctoral Fellow, Department of Building Technology, Linnaeus University, Sweden.
Email: osama.abdeljaber@lnu.se
[3] Professor, Department of Electrical Engineering, Qatar University, Qatar.
Email: mkiranyaz@qu.edu.qa
[4] Associate Professor, Department of Civil Engineering, Qatar University, Qatar.
Email: mhussein@qu.edu.qa
[5] Professor, Department of Signal Processing, Tampere University of Technology, Finland;
Email: moncef.gabbouj@tut.fi
[6] Professor, Department of Aerospace Engineering, University of Michigan, Ann Arbor, MI, USA.
Email: daninman@umich.edu

**Abstract**

Monitoring structural damage is extremely important for sustaining and preserving the service life of civil structures. While successful monitoring provides resolute and staunch information on the health, serviceability, integrity and safety of structures; maintaining continuous performance of a structure depends highly on monitoring the occurrence, formation and propagation of damage. Damage may accumulate on structures due to different environmental and human-induced factors. Numerous monitoring and detection approaches have been developed to provide practical means for early warning against structural damage or any type of anomaly. Considerable effort has been put into vibration-based methods, which utilize the vibration response of the monitored structure to assess its condition and identify structural damage. Meanwhile, with emerging computing power and sensing technology in the last decade, Machine Learning (ML) and especially Deep Learning (DL) algorithms have become more feasible and extensively used in vibration-based structural damage detection with elegant performance and often with rigorous accuracy. While there have been multiple review studies published on vibration-based structural damage detection, there has not been a study where the transition from traditional methods to ML and DL methods are described and discussed. This paper aims to fulfill this gap by presenting the highlights of the traditional methods and provide a comprehensive review of the most recent applications of ML and DL algorithms utilized for vibration-based structural damage detection in civil structures.

**Keywords:** Structural damage detection • Structural Health Monitoring • Vibration-based methods • Machine Learning • Deep Learning • Infrastructure health • Artificial Neural Networks • Civil infrastructure

## 1. Introduction

Structural damage is intrinsic in engineering structures and it is predominantly prone to propagate due to various environmental and mechanical factors. The short term and long term damages cause the structures to age and shorten the design life which makes the monitoring process an important aspect for structures [1]. Monitoring structural damage on structures started with visual inspection and evolved in time with tremendous developments in Structural Health Monitoring (SHM) and Structural Damage Detection (SDD) fields. A large number of techniques have been developed to detect, localize and quantify structural damage in an attempt to make the monitoring process more feasible [2–4]. Numerous vibration-based damage detection techniques have been studied where the vibration response of the monitored structure is recorded and analyzed to assess structural damage and make decisions on the structural health [5]. Consequently, different techniques on vibration-based SDD have been researched, proved successful and articulately accepted over the decades [6].

Damage is traditionally defined as a change in the geometric or material characteristics of a system that adversely affects its performance, safety, reliability, and operational life [7,8]. According to this definition, damage does not always indicate a complete failure of a system, yet a comparative deterioration of the system functionality causing a suboptimal performance [9–11]. If no remedial action is taken, damage may accumulate until reaching the failure state. Systems may fail in a gradual or sudden manner depending on the type of the damage [12,13]. For instance, failure due to corrosion or fatigue usually occur over long time periods, while earthquakes and fire-induced damage can lead to a rapid failure [14].

Engineering structures are susceptible to human-induced and environmental factors which speed up the accumulation and propagation of damage and shorten the structural life cycle [15]. In an attempt to monitor the structural health, SHM systems have been implemented in mechanical, aerospace, and civil engineering applications [16–18]. SHM is a broad and highly interdisciplinary research field that involves experimental testing, system identification, data acquisition and management, as well as long-term measurement of environmental and operational conditions [19–22]. The most critical component of SHM is damage detection, which is defined as a systematic and automatic process of identifying the existence of a damage, and then localizing and assessing the severity of it (quantification) [23].

A typical damage detection system consists of software and hardware components, as shown in Figure 1. The hardware part is composed of the sensing and data acquisition interface used to collect measurements which may usually include accelerometers, velocimeters, strain-gauges, load cells, or fiber optic sensors along with data acquisition modules [24,25]. The software component of a damage detection system is an arsenal of signal-processing and pattern recognition algorithms designed to translate the signals acquired by the sensing interface into essential information that reflects the condition of the structure being monitored [26,27].

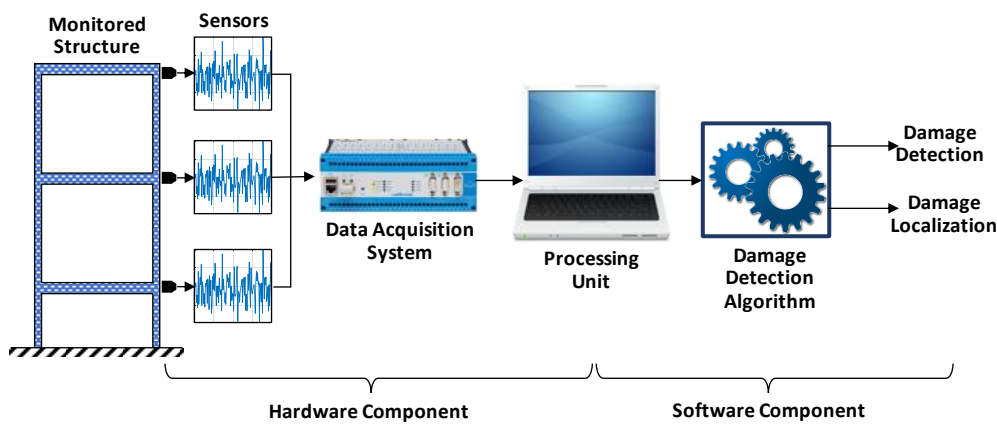

Figure 1 – Main components of structural damage detection systems.

Numerous applications of damage detection can be found in almost all areas of engineering research. The most successful and well-established damage detection systems are those used in mechanical engineering for vibration-based condition monitoring of rotating machines and machine parts [28–30]. Such systems utilize the vibration response of a machine recorded in terms of displacement, velocity, or acceleration to evaluate the condition of its bearings, gears, and shafts [31]. Damage detection is processed by extracting a "vibration signature" from the response spectrum and then applying a pattern recognition method to compare the current signature to that of the reference (i.e. undamaged) condition [32–34]. The high efficiency of vibration-based SDD systems via SHM of rotating machines [35] is based on the fact that the vibration response of such machines is minimally affected by the operational and environmental conditions [36–38]. This means that any notable difference in the vibration signature of a system can be attributed to a certain type of damage. Since the types and locations of damage in rotating machines are well-defined, it is quite feasible to correlate each change in the vibration signature to a specific type and location of damage [39].

Civil structures are constantly exposed to human-induced and environmental factors which shorten their life cycle [40–42]. Damage is commonly encountered in civil infrastructure due to creep, corrosion, shrinkage, fatigue and scour [43–45]. Hence, it is of utmost importance to continually monitor civil structures to evaluate their structural condition and provide early warning against structural damage [46]. The traditional approaches for damage diagnosis of civil structures are mainly based on visual inspections. Yet, there are several issues that hinder the application of these methods in practice [47]. Primarily, the size of civil structures is relatively large, which means that regular inspection can be both laborious and time consuming. Second, since the traditional approaches depend on human judgment, they definitely demand skilled and highly-trained labor. Third, in most of the cases, it is difficult to visually inspect the load-carrying structural members (i.e. footings, columns, beams, and slabs) as they are usually covered by non-structural components and decorative coverings such as flooring, cladding, false ceiling, etc. Nevertheless, the success of vibration-based methods in condition monitoring of machinery has motivated researchers to implement similar techniques in SHM of civil infrastructure. As a result, a wide variety of vibration-based SDD systems has been developed [48,49]. The ultimate goal of these systems is to overcome the issues associated with the traditional SDD approaches by providing a systematic, feasible and consistent way of identifying the presence, locating, and quantifying the severity of the structural damage based on the vibration response of the monitored structure [50–54].

Meanwhile, plenteous improvements in computational power and advancements in chip and sensor technology have enabled the use of Machine Learning (ML) techniques in engineering applications. While ML is the science of developing intelligent algorithms capable of acquiring knowledge automatically from the available data, the objective of ML algorithms is to provide machines with a human-like ability of learning-by-example and pattern recognition. Recent investigations have shown that ML approaches are more meticulous and superior than the traditional rule-based methods especially in dealing with problems where the data is fuzzy [55], vague, or noise-contaminated [56]. For example, in structural engineering, ML has been applied for structural system identification [57], seismic performance assessment [58], modeling of compressive [59] and tensile [60] concrete strength, shear capacity estimation of FRP-reinforced concrete beams [61], vibration control [62], as well as structural health monitoring [63]. A general overview of ML utilization in structural engineering applications is presented in [64].

During the last decade, different ML tools have been utilized to develop a wide array of parametric and nonparametric vibration-based SDD systems for civil structures. These tools include but are not limited to Artificial Neural Networks (ANNs), Support Vector Machines (SVMs), Self-Organizing Maps (SOMs), Convolutional Neural Networks (CNNs), etc. Extensive analytical and experimental studies have been conducted in an attempt to demonstrate the efficiency of these SDD systems. While there has been multiple review studies on vibration-based SDD in the literature [65–73], there has not been a study where the transition from traditional methods to recent Machine Learning (ML) and Deep Learning (DL) paradigms are described and discussed. To be more specific, [66] only covers parametric ML-based SDD methods, while this paper reviews both parametric and nonparametric methods. On another note, [67–70,72] focus predominantly on traditional vibration-based SDD methods without highlighting ML and DL methods while this paper covers work utilizing ML and DL based methods. While [71] highlighted both traditional and ML methods, it was published over 13 years ago during which tremendous improvements have been accomplished. It is also noted that even though [65,73] are providing reviews on SDD, they weren't intended and organized to serve as review papers. As such, this article aims to fulfill the gap by presenting the highlights of the traditional methods and provide a broad review of the most recent applications of ML and DL algorithms for vibration-based SDD in civil structures.

The rest of the paper is organized as follows. A general background on AI, ML and DL is presented in Section 2. The categorization of automated parametric and nonparametric vibration-based SDD methods is discussed and reviewed in Section 3. In Section 4, the applications of conventional ML algorithms in both parametric and nonparametric vibration-based SDD methods are reviewed in detail. The recent applications of DL methods utilized for vibration-based SDD in civil structures are reviewed in Section 5. Conclusions are drawn and future study topics are discussed in Section 6.

## 1.1 Methodology

For this paper, the authors reviewed a total of 112 journal and conference articles related to vibration-based structural damage detection including 58 articles on traditional methods and 54 articles on ML and DL methods. The methodology for selecting these articles can be summarized as follows:

- The articles reviewed in this paper were collected from well-known databases including ASCE library, IEEE Xplore Digital Library, Scopus, Web of Science, Science Direct, Sage, and Wiley Online Library.
- The literature search was conducted using keywords such as "structural damage detection", "vibration-based damage detection", "global structural damage detection", "non-parametric damage detection", "parametric damage detection", "neural networks for damage detection", "machine learning structural vibration", etc.
- All peer-reviewed journal articles related to ML and DL applications on vibration-based damage detection in civil structures and published between 1997 and 2019 were selected. Studies on local structural damage techniques (e.g. impedance-based methods) were excluded since they are beyond the scope of this review paper. Additionally, articles related to vibration-based damage detection techniques in mechanical and aerospace structures were filtered out.

- Relevant conference papers from prominent conferences such as the International Conference of Sound and Vibration (ICSV), International Modal Analysis Conference (IMAC) and ASCE Geotechnical and Structural Engineering Conference were also selected and reviewed.

In the light of the above, Table 1 summarizes the name and number of the journal and conference papers utilizing ML and DL tools in vibration based SDD that are visited and reviewed for this publication. In addition, the publications shown in Table 1 are grouped in bar chart format in Figure 2 based on the publication year. It is observed in Figure 2 that the amount of work regarding the ML and DL tools in vibration based SDD has been in increasing trend over the last 15 years which brings the need for a new review paper to discuss the improvements and state-of-the-art in the field.

Table 1 - Journal and Conference Papers utilizing ML and DL tools in Vibration-Based Structural Damage Detection

| Journal or Conference | Number of Articles |
| --- | --- |
| Engineering Structures | 6 |
| Journal of Sound and Vibration | 5 |
| Structural Health Monitoring | 4 |
| Advances in Structural Engineering | 3 |
| International Conference on Noise and Vibration Engineering | 3 |
| Australian Journal of Structural Engineering | 2 |
| Expert Systems with Applications | 2 |
| International Modal Analysis Conference (IMAC) | 2 |
| Journal of Intelligent Material Systems and Structures | 2 |
| Shock and Vibration | 2 |
| Structural Control and Health Monitoring | 2 |
| Applied Soft Computing | 1 |
| Cluster Computing | 1 |
| Composite structures | 1 |
| Computers and Structures | 1 |
| Conference Proceedings of the Society for Experimental Mechanics Series | 1 |
| Engineering Applications of Artificial Intelligence | 1 |
| Journal of Architectural Engineering | 1 |
| Journal of Bridge Engineering | 1 |
| Journal of Engineering Mechanics | 1 |
| Journal of Performance of Constructed Facilities | 1 |
| Journal of Theoretical and Applied Mechanics | 1 |
| Jurnal Teknologi | 1 |
| KSCE Journal of Civil Engineering | 1 |
| Mathematical Problems in Engineering | 1 |
| Measurement | 1 |
| Meccanica | 1 |
| Mechanical Systems and Signal Processing | 1 |
| Mechanics of Advanced Materials and Structures | 1 |
| Neural Networks | 1 |
| Neurocomputing | 1 |
| Structural Design of Tall and Special Buildings | 1 |

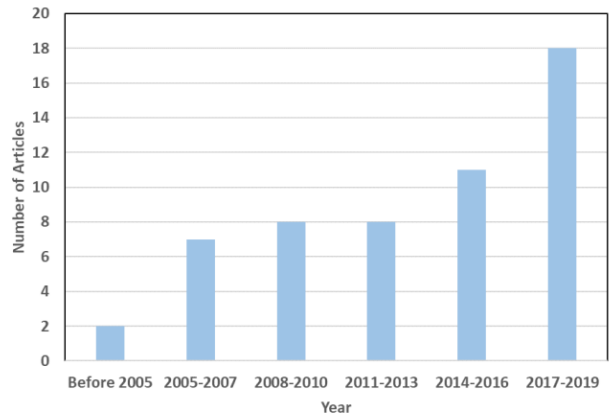

Figure 2 - Number of papers related to ML and DL applications in structural damage detection published between 1997 and 2019.

## 2. Artificial Intelligence, Machine Learning, and Deep Learning

Artificial Intelligence (AI) is a field of computer science aiming at developing machines that exhibit human-like intelligence in solving problems and performing tasks. The early applications of AI particularly targeted rule-based problems that are intellectually challenging for humans yet straightforward for computers. The knowledge required to solve such problems could be easily conveyed to a computer as a rigid list of hand-coded "if" and "else" statements designed by a human expert [74]. This knowledge-based approach to AI enabled machines to surpass human ability in abstract and formal tasks that are governed by a relatively brief set of rules such as playing chess [75]. Nevertheless, knowledge-based AI systems have dramatically failed in "everyday" tasks that appear to be automatic and straightforward to an average human being such as recognizing faces, detecting objects, and understanding speech. This is because such intuitive tasks require considerable experience and informal awareness about the world that cannot be translated into an explicit list of formal rules. Therefore, a major challenge faced by modern AI systems was to find alternative ways of teaching intuitive and common-sense knowledge to computers [76].

For overcoming the drawbacks of knowledge-based approach, the concept of ML was introduced to AI systems. ML is the ability of a machine to acquire its own knowledge automatically from the data [77]. Hence, ML algorithms allow computers to "learn" the knowledge required for carrying out a specific task by analyzing a sufficient number of relevant data samples in a systematic manner. Basic ML algorithms require the data representation in terms of a fixed number of features. Therefore, before applying the algorithm, it is pertinent to preprocess data sample to extract the certain features or attributes that represent the most characteristic pieces of information. This process is referred to as "feature extraction" [78]. Next, the processed data samples, which are expressed in extracted feature terms, are used to train the ML system based on a particular ML training method in order to learn how these features correlate to discriminate different data patterns. The relationships among knowledge-based AI, ML, DL approaches are illustrated in Figure 3(a); meanwhile a comparison between these approaches is depicted in Figure 3(b).

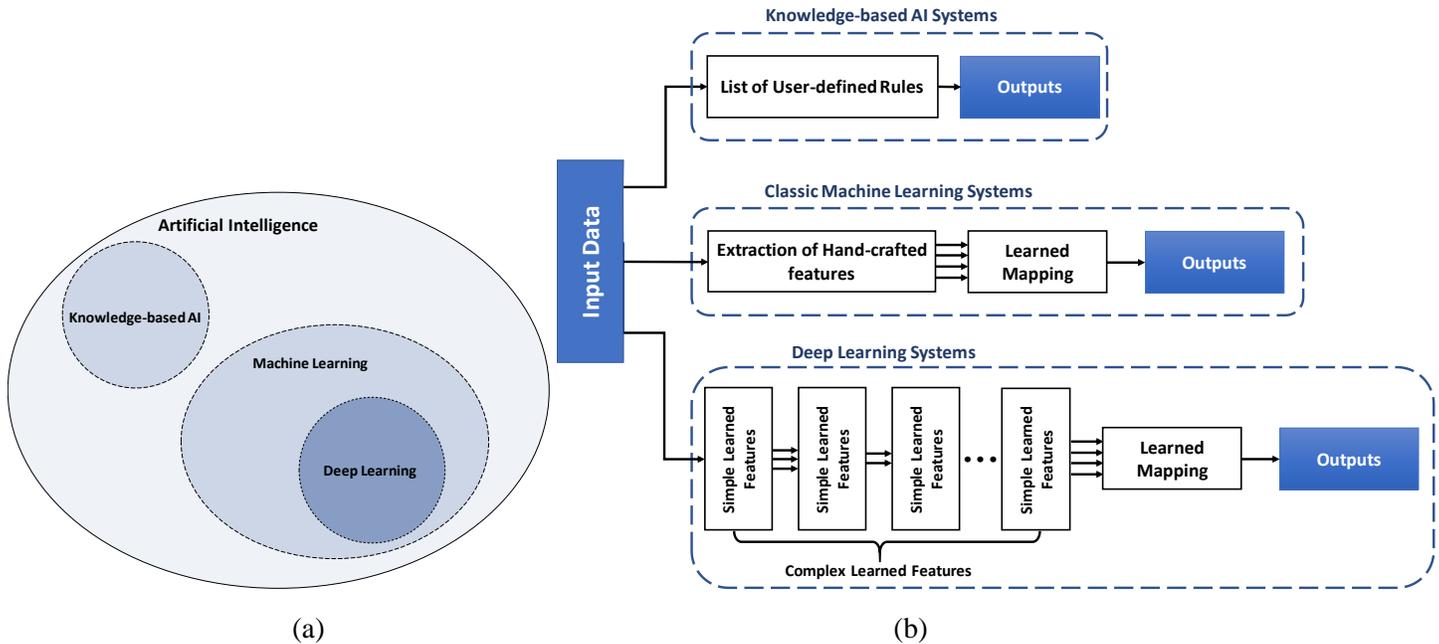

Figure 3 – (a) Venn diagram presenting the relationship among different AI systems; (b) Knowledge-based, Machine-Learning, and Deep Learning systems.

## 2.1 Machine Learning

As an application of AI, ML basically provides systems the ability to automatically learn and improve from experience without being explicitly programmed [79]. For clarification and better understanding of ML, following example can be considered. Suppose the goal is to develop an AI medical diagnosis method for predicting whether or not the patient suffers from a certain type of cancer. The data available for training the system consists of a large number of samples. Each data sample consists of an input, which is a record of clinical observations obtained for a previous patient, and an output, which is a binary variable representing weather or not this patient was diagnosed with the cancer. The first step toward training the system would be to ask a medical doctor to list the symptoms that are likely to be associated with the targeted type of cancer. The training dataset can then be analyzed to extract information regarding these symptoms from each patient record (i.e. feature extraction process). Next, the processed dataset, which consists of information regarding the listed symptoms extracted from each patient record along with the associated outputs (i.e. patient being diagnosed with cancer or not), is used to train the system according to a specific ML algorithm. Through the training process, the system will automatically learn the relationships between the symptom patterns and the output, i.e., cancer (malignant) or normal (benign). After verification tests, the trained system can be used to diagnose any future patient given the medical record in terms of the listed symptoms.

ML algorithms can generally be classified into unsupervised and supervised algorithms [80]. Per the aforementioned example, supervised algorithms require a dataset consisting of human-labeled data for training. As such, the primary purpose of the supervised learning is to discover the optimal mapping from the inputs to the desired (or target) outputs [81]. Therefore, supervised algorithms require a human "supervisor" to assign each data sample by a correct label or target before running the training [82]. Unsupervised learning algorithms, on the other hand, require input-only data without any labeling. The objective of unsupervised learning is to investigate the distribution of the data in order to obtain useful information regarding its underlying structure [83]. Based on the above, the tasks performed by ML systems can be summarized as the following:

1. *Classification:* The objective of this task is to determine which category the input belongs to. The medical diagnosis system discussed earlier serves as an example of classification task since the input to the system is classified into either "cancer" or "no cancer" categories.
2. *Regression:* In this task, the goal is to model the relationship between a numerical output and a number of inputs. The only difference between regression and classification is the format of the output.
3. *Prediction:* Prediction is a special type of regression in which the objective is to foresee the future values of a given time series.

4. *Clustering:* The target of clustering is to divide the input dataset into clusters with similar examples [84]. Unlike classification, regression, and prediction tasks which are performed using supervised methods, clustering is conducted in an unsupervised manner such as self-organizing maps.

**2.2 Deep Learning**

As discussed earlier, basic supervised and unsupervised ML algorithms require feature extraction in advance to represent the input data in terms of a fixed number of hand-crafted features. While the performance of such algorithms relies significantly on the choice of extracted features, it is critical to select the right group of features optimally representing the most distinctive and applicable properties of the input data [85]. As a matter of fact, after extracting the features, it becomes relatively easy for a simple ML algorithm to obtain the mapping between the extracted features and the desired output. This approach to ML is useful in the cases where it is possible to manually determine the extracted features like in the previous example where the features (i.e. the list of symptoms) could be decided by a human expert.

Nevertheless, for numerous tasks, it is not very easy to manually select a good group of features to be used for training the AI system [86]. For better understanding, a computer vision system for vehicle detection in images would be an appropriate example to consider. The data available to train the system consists of a number of input-output samples. For each case, the system input is an image of traffic on a street, while the output is the actual locations of vehicles in that image. Before training the system using a conventional ML algorithm, it is necessary to represent each input image in terms of a set of features that are expected to give strong clues regarding the location of vehicles in the input image. However, this may not be feasible outcome since the appearance of vehicles in photographs significantly depends on factors like the camera position; viewing angle; light and shadow conditions; occlusion with obstacles and other vehicles, etc. Therefore, while those manually selected (or hand-crafted) features may work for certain cases, they may entirely fail for others.

Based on the above, in order to avoid the hand-crafted features in complex ML applications, Deep Learning (Deep Neural Learning or Deep Neural Network) methods have been introduced. DL is indeed a subset of ML within AI context that has networks capable of learning unsupervised from unstructured data. DL, also known as representation learning, are a special type of ML methods capable of extracting the optimal input representation directly from the raw data without user intervention. In other words, DL algorithms can learn not only to correlate the features to the desired output, but also to carry out the feature extraction process itself. Therefore, a DL system with proper training can indeed find the direct mapping from the raw inputs (e.g. the images in the previous example) to the final outputs without the need to extract features in advance [87]. DL is then able to explain high-level and abstract features as a hierarchy of simple and low-level learned features [74]. This ability allows DL algorithms to deal with complex tasks by breaking them down into a large number of simple problems. Recent studies have revealed that relying on learned features instead of hand-crafted ones results in much better performance in challenging tasks such as object detection [88], image classification [89], and classification of electrocardiogram (ECG) beats [90].

**3. Vibration-based structural damage detection methods that are not based on Machine-Learning**

SDD techniques are predominantly classified into global and local methods. While all vibration-based SDD methods are traditionally considered as global methods; the local SDD methods are based on detecting and quantifying structural damage in a relatively smaller scale without focusing on and using the vibration response of structures. As the detection range for local methods is relatively small, most of the methods in Non-Destructive Testing and Evaluation (NDTE) are mostly considered as local methods [91,92]. Tools such as Ultrasonic Testing (UT), Acoustic Emissions (AE), Infrared Thermography (IRT), Radiographic Testing (RT), Magnetic Flux Leakage (MFL), Magnetic Particle Testing (MT), Digital Image Correlation (DIC), Liquid Penetrant Testing (PT), Laser Testing Methods (LM), Ground Penetrating Radar (GPR), Leak Testing (LT), Visual Testing (VT) and numerous optical methods are utilized for inspecting, testing and evaluating of structural components and assemblies at local areas of infrastructure [93–98]. For instance, electro-mechanical impedance-based damage detection approaches [99] are commonly used for local damage detection in small structural members such as thin plates and beams. The basic idea of impedance-based methods is to attach a number of piezoelectric patches to the monitored structural member. The patches, which operate as both actuators and sensors, are driven by a very high frequency excitation (typically higher than 30 kHz). The electro-mechanical impedance signal across the patch terminal is then measured and used as a damage indicator that reflects the local condition of the area very close to the location of the patch. Even though the NDTE methods and other local methods are also used in damage identification, localization and quantification; the authors did not include the review of these methods in this manuscript since the manuscript is solely focused on vibration-based SDD. Nevertheless, it is crucial to note that such local techniques alone are not sufficient for SHM of large-scale civil structures. As such, for complete understanding of the structural condition of a large structure, an efficient SHM system must combine local and global damage detection techniques.

SHM and SDD through vibration response of structures has been an area of research over the decades. Researchers focused on the time, frequency and modal domains searching for the presence, location and the severity of damage on engineering structures [100–103]. Various techniques are introduced and numerous algorithms are developed for vibration-based SDD in civil, mechanical and aerospace engineering. Global (i.e. vibration-based) SDD methods utilize the vibrations response of the monitored structure to understand the overall condition of the structure [104–106]. Such methods require a network of accelerometers placed on strategic points of the monitored structure in order to record its vibration response. Specialized algorithms are then used to translate the measured acceleration signals into indices that reflect the presence of damage, and localize and quantify the severity of the damage [107,108]. The concept of vibration-based damage diagnosis was reported to be pronounced in 1800's, when railway workers used to assess the condition of train wheels qualitatively by tapping them with a hammer and listening to the resulting sound [109]. However, the development of quantitative approaches of global damage detection was only made possible in the 1980s due to the evolution of computing and sensing technology. Simultaneous advancements in computer chip technology enabled the use of faster processors resulting in more detailed finite element (FE) models, and quicker and more efficient model updating of structures [110]. Compared to the local methods, vibration-based SDD techniques offer the following advantages [1]:

1. Vibration-based methods do not require a network of closely-spaced sensors. A limited number of accelerometers is usually sufficient for identifying the dynamic properties even in very large and complex structures.
2. To apply a vibration-based method, it is not required to know the expected location of damage in advance.
3. The instruments required for running most vibration-based methods can be easily carried, handled, and attached to the monitored structure.

Depending on the extracted information from the measured signals, vibration-based SDD techniques can be further categorized into nonparametric and parametric methods as shown in Figure 4. These two types of vibration-based methods are reviewed in the next two subsections.

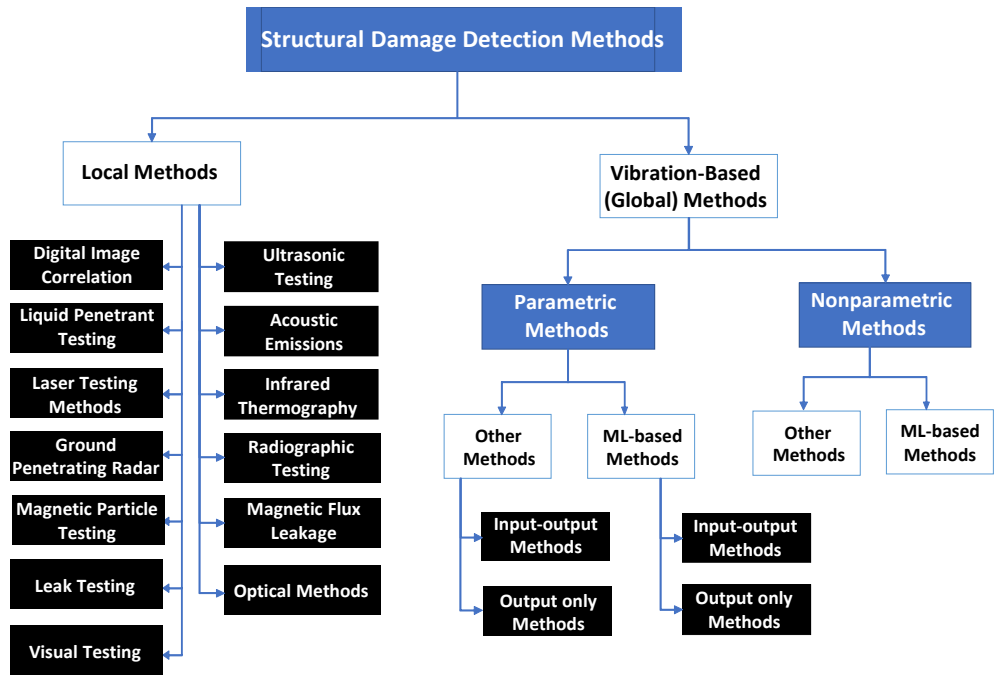

Figure 4 - Classification of structural damage detection methods.

## 3.1 Parametric vibration-based SDD methods

The meaning of the word "parametric" as an adjective is *"relating to or expressed in terms of a parameter or parameters"*. In parallel to this definition, in parametric vibration-based SDD methods, vibration signals acquired by sensing interface are utilized to determine the unknown dynamic parameters of the structural system. These parameters are simply the physical properties of the structure such as modal frequencies, modal mass, modal damping, stiffness, and mode shapes [111,112]. Parametric methods attempt to detect structural damage by comparing the undamaged dynamic parameters of the structure to the damaged parameters. As such, the changes observed in the dynamic parameters with respect to a predefined reference condition (datum) can simply be used as indicators of the formation, location and severity of structural damage [113]. While the earlier parametric methods have relied basically on correlating structural damage to the changes in the modal characteristics, the idea was to apply a known input excitation on the undamaged structure and measure the vibration response using an array of accelerometers (as shown in Figure 5). The artificial excitation can be applied using a modal sledge hammer or an electro-dynamic shaker. The input/output signals are then processed by a modal identification algorithm to determine the modal parameters of the reference condition. For evaluating the integrity of the structure after a certain amount of time, the same exercise is conducted again, and the extracted modal parameters are compared to those of the reference condition.

Several time-domain algorithms input-output modal identification techniques have been utilized for parametric damage detection including but not limited to Complex Exponential Analysis (CEA) [114], damped CEA [115], and poly-reference time domain method (PTD) [116]. Also, various frequency-domain algorithms have been utilized for the same purpose like simple peak-picking of the natural frequencies from frequency response functions [117], Frequency Domain Decomposition (FDD) [118], and Complex Mode Indicator Function (CMIF) [119]. A specific review of input/output identification techniques can be found in [72]. Parametric methods that depend on input/output modal methods were relatively successful in mechanical and small aerospace structures; however, it is relatively difficult to implement them in SHM of civil structures since it is usually impractical to excite an infrastructure system artificially with a known input. A hammer or shaker excitation would be almost negligible for a heavy and robust civil structure. As a solution, researchers have investigated several output-only modal identification techniques for modal parameter estimation per ambient vibration response (depicted in Figure 6). Different output-only approaches have been utilized in the context of SDD including peak-picking of natural frequencies [120] from auto and cross-spectral densities, random decrement (RD) method with FDD, and Stochastic Subspace Identification (SSI) procedure [121].

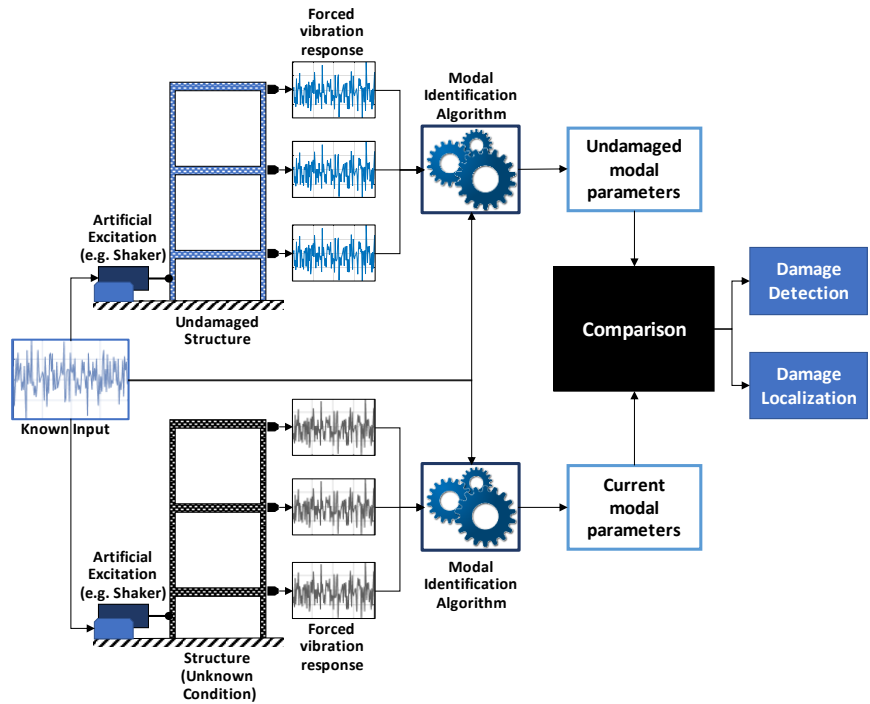

Figure 5 - A schematic of input-output parametric vibration-based damage detection methods.

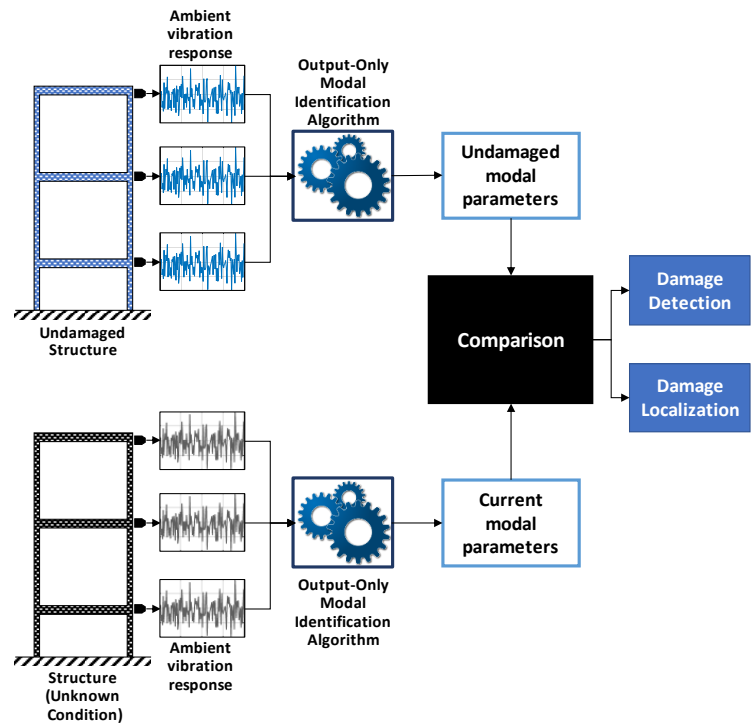

Figure 6 - A schematic of output-only parametric vibration-based damage detection methods.

As an example for output-only parametric damage detection, Gul et al. [119] proposed Complex Mode Indicator Functions (CMIF) combined with Random Decrement (RD) technique to synthesize the unscaled modal flexibility matrix per the output-only recordings, assuming that the input excitation is unknown. The extracted flexibility matrix was utilized to estimate the deflection profile of the monitored structure. Experimental tests on a two-span laboratory grid structure (5.5 m x 1.8 m plan dimensions) showed that the deflection profile can be used to indicate changes in the structural boundary conditions. As such, the modal parameters obtained from the ambient vibrations enabled the unscaled flexibility matrices distinguish the damaged condition and its location. The unscaled flexibilities obtained from undamaged and damaged conditions are shown in Figure 7. Based on this figure, the deflection at the span where the corner joints are restrained is found to be less than the deflections at the neighboring span where all the joints are unrestrained. The comparison of two unscaled deflected shapes shows that there has been a modification in the structural system while it provides spatial information about the location of the damage. As a result, the authors showed that the dynamic properties identified with ambient vibration testing successfully correlate with impact testing and the finite element (FE) model results. After the laboratory work, when the same technique is utilized to successfully identify the modal properties of a long-span bridge, the accuracy of this output-only parametric damage detection method is verified. The fact that the proposed method succeeded in both small-scale and full-scale structures, the methodology can be generalized and used in other applications.

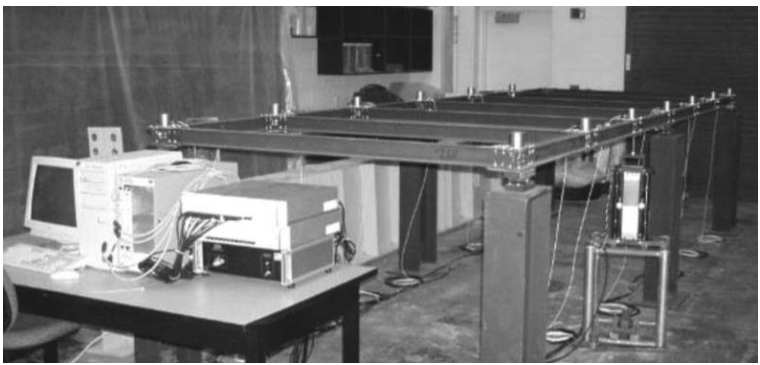 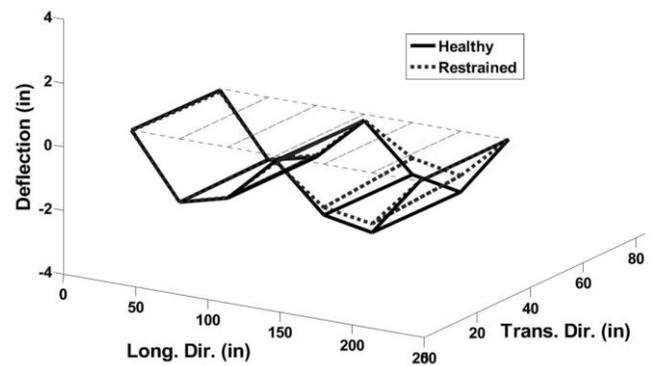

Figure 7 – The laboratory structure of Gul et al. [119] study (left). The unscaled flexibilities obtained from damaged and undamaged structures (right).

Similarly, Frizzarin et al. [122] used RD signature method to identify the free vibration response of concrete structures from the output-only vibration signals measured under ambient conditions. The computed free vibration response was then utilized to determine the damping ratio of the monitored structure. The RD signature is a free decay and can be expressed as follows:

$$z(\tau) = \frac{1}{N}\sum_{k=1}^{N} y(t_k + \tau) \tag{1}$$

where, $z(\tau)$ is the RD signature, $\tau$ is the time reference of the sub-segments, $N$ is the number of averages, $y$ is the recorded data, and $t_k$ is the time at which the triggering level is crossed. Experimental results obtained by testing a large-scale concrete structure suggested that variations in modal damping ratio values could also be utilized as damage indicators. It was reported that the damping ratio increases per the seismic damage. It is also reported that there is an evident correlation between the nonlinear damping ratio and the stiffness reduction per the severity of seismic damage on the structure. The four damage identification methods in this study were all observed to be in good agreement and the nonlinear damping parameter was found to be a reliable damage index. Considering 1.0 for the undamaged condition stiffness, the stiffness values identified by the three different methods all decrease consistently as the seismic damage becomes more severe (Figure 8). The fact that the nonlinear damping-based method was proved to be successful for seismic damage of different levels while remaining consistent with the observed damage, the baseline-free feature of this method provides a great advantage. All these features enable the approach to be generalized and used in other applications.

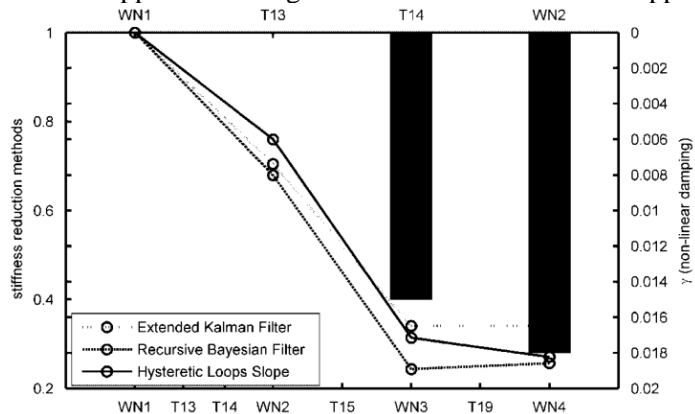

Figure 8 – Comparison of damage identification methods in Frizzarin et al. [122].

Another example is the method introduced by Yin et al. [123] which utilized Natural eXcitation Technique (NeXT) along with Eigenvalue Realization Algorithm (ERA) to identify the modal characteristics directly from the ambient response of transmission towers. One of the goals of the method was to extend the dynamic reduction method to SDD of transmission towers which are typically statically indeterminate structures. Therefore, the modal properties corresponding to the undamaged condition were utilized for detecting and quantifying damage based on the dynamic reduction method. After modal parameter identification, focusing on the damage on secondary structural members, the proposed NeXT-ERA technique utilizes a small number of sensors to detect sub-structure damage by calculating equivalent stiffness loss. This method essentially differentiates the SDD task into solving a group of nonlinear equations. The proposed parametric damage detection approach was verified numerically using a FE computer model of the transmission tower. Using information from different mode combinations, the identified damages with modeling error are shown in Figure 9. While the authors indicated that the verification is successfully done per simulated noisy data from a 3D model under both single and multiple damage cases, the experimental verification of the proposed method was pointed out as a future work in the conclusions chapter of the paper. Therefore, the generalization of the method for other structures has not been explored yet. Based on the success of the method for a 3D FE model of a spatial structure, the method is promising especially for the planned use of force transducers to measure the force on transmission cables as discussed in the future works chapter.

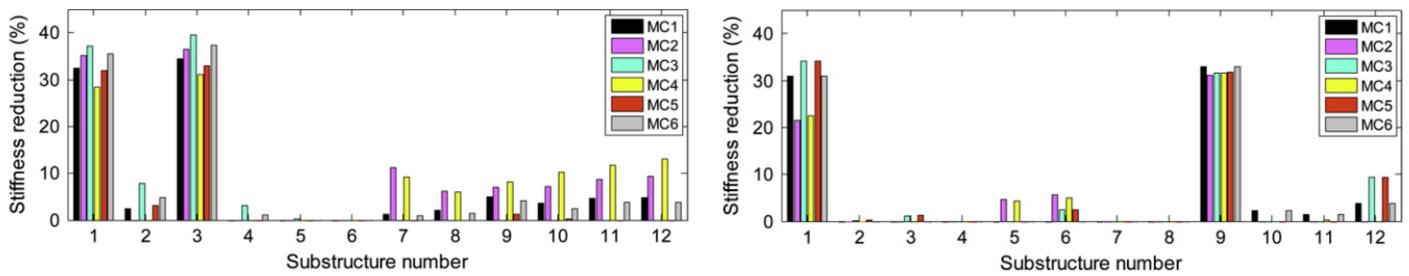

Figure 9 - Identified damages with modeling error using different mode combinations: Case 2 (left) and Case 3 (right) in Yin et al. [123].

In another related SDD study, Miguel et al. [124] developed an output-only parametric method by integrating Stochastic System Identification (SSI) modal estimation technique with a harmony search algorithm. The method was tested both numerically through FE simulations and experimentally using three small cantilever beams. The function for damage quantification is Eq. (2):

$$\Pi(\alpha) = \|[F]_E - [F(\alpha)_E]\|^2_{Fro} \qquad (2)$$

where $\alpha$ is the vector depicting the damage parameters, $\|\ \|_{Fro}$ is Frobenius norm for the residual matrix, $F_E$ indicates the modal flexibility matrix from the identified results, and $F_A$ is the modal flexibility matrix calculated from the analytical model with vector damage parameters. The results revealed the ability of the proposed method to identify damage cases introduced by cutting the beams at five different depths (2.5 mm, 3.5 mm, 5mm, 8.5 mm and 16 mm). The cuts meant a reduction in moment of inertia and stiffness at the cut locations. Vibration response was recorded at every 0.3m of the test beams and stiffness reduction factors are plotted for each damage case (Figure 10). The first five bending mode frequencies were determined for pre-damage and post-damage conditions. It was reported that the levels of SDD and localization was acceptable for both excitations and noise level; however, the damage detection and localization were relatively more accurate for small beam cuts than deeper cuts. The identified damage patterns are experimentally consistent for all damage locations and severities; however, the methodology needs to be verified on a large scale structure before the approach can be generalized and used for other damage assessment studies.

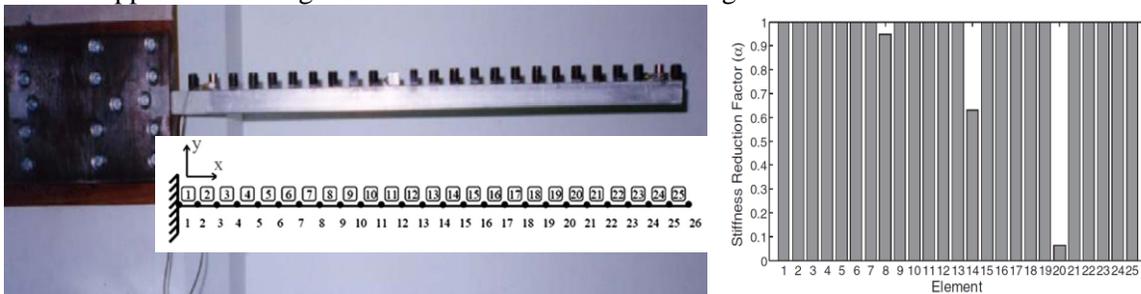

Figure 10 – Experimental setup (left) and the stiffness reduction factor for one damage location (right) in Miguel et al. [124].

All of the aforementioned parametric SDD methods assume that a structural damage significantly affects the mass, bending stiffness and/or modal damping of structures. While this assumption sounds intuitive in general, for the practical field applications of parametric methods, the following shortcomings are identified [12,119]:

1. Recent investigations demonstrated that certain types of structural damage cannot always be associated with changes in global dynamic characteristics especially those corresponding to the lower modes of vibration. Local structural damage only affects higher frequency modes which are typically difficult to identify using output-only methods.
2. Modal parameters are affected by factors other than damage such as temperature, moisture, and measurement noise. Therefore, changes in these parameters do not necessarily indicate actual structural damage.
3. Output-only parametric methods rely on sophisticated system identification algorithms to solve the inverse problems on modal parameter estimation per the ambient structural response. Implementation of such algorithms is typically reported to be infeasible and therefore unsuitable for real-time damage detection applications.
4. Parametric damage detection methods are centralized, which means that all signals should be transferred to a central processing unit before carrying out the damage identification process. Yet, transferring and synchronization of large amounts of measurements is particularly problematic when using wireless sensor networks. As such, centralized methods are not robust against sensor failure since they require all sensors in the network to be fully functional.

## 3.2 Non-parametric vibration-based SDD methods

After the primary use of parametric methods in vibration-based SDD, the focus shifted to nonparametric damage detection methods in time. Unlike parametric approaches which rely on system identification, non-parametric SDD methods utilize statistical means to detect structural damage directly from the measured accelerations [125]. Nonparametric methods are able to extract damage features that cannot be easily attributed to physical changes of the structure. As illustrated in Figure 11, non-parametric methods combine time series modeling with statistical classification. The first step is essentially extracting damage-sensitive features from raw signals using a **time series modeling technique**. The extracted features are then processed either by a classifier or by an outlier detector to assess the current health state of the structure [126,127].

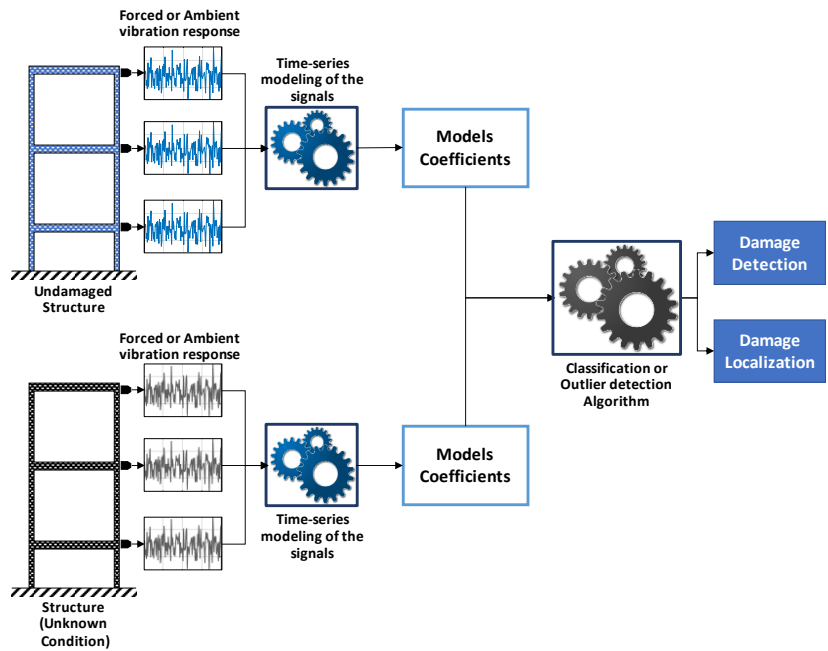

Figure 11 - A schematic of nonparametric vibration-based damage detection methods.

Several nonparametric vibration-based SDD methods are published in the literature. For instance, Nair and Kiremidjian [128] proposed an SDD approach which uses Auto-Regressive Moving Average (ARMA) for feature extraction along with a Gaussian Mixture Model (GMM). Mahalanobis distance between the extracted feature vector and the features corresponding to the undamaged condition was utilized for SDD. Per the damage extent shown in Figure 12, the Mahalanobis distance (DM) tend to increase for damage patterns 6, 3, 4, 5, 1, and 2, which is consistent with the increasing level of damage cases used in the benchmark structure [129] adopted in the study. While this technique is verified against simulated vibrations obtained from a benchmark structural health monitoring problem, it needs to be noted that the algorithm is valid for linear stationary signals. In this study, it is recommended that the initial testing of the algorithm and further investigations are needed to test the validity of the procedure with other data and feature vectors [128].

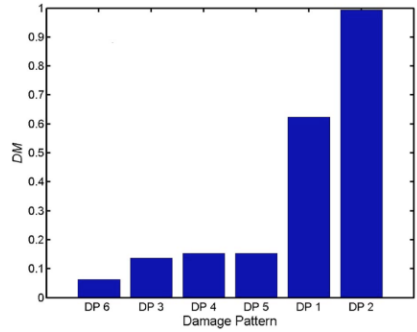

Figure 12 – Mahalanobis distance (DM) with respect to damage patterns of the structure in Nair and Kiremidjian [128].

A similar approach was developed by Carden and Brownjohn [130] who used features extracted by ARMA approach together with an unsupervised classifier. While the paper is making use of time series response analysis to provide a statistical method of inferring structural change, the classification of features was performed on the basis of the summation of the square of residuals. The linear stochastic models of structural time response histories using Autoregressive Moving Average (ARMA) models can be built for randomly excited structures. An ARMA model of order *(p, q)* can be formulated per Eqs. (3-5).

$$\phi(B)z_t = \theta(B)a_t \tag{3}$$

$$\phi(B) = 1 - \phi_1 B - \cdots - \phi_p B^p \tag{4}$$

$$\theta(B) = 1 - \theta_1 B - \cdots - \theta_q B^q \tag{5}$$

where $\phi(B)$ is the AR function of order $p$; $\theta(B)$ is the MA function of order $q$; $z_t$ is the time history response of the structure; and the $a_t$ is the series of Gaussian distributed random shocks acting on the structure.

The study used experimental data collected from three different structures. Experimental data recorded under forced vibration tests were conducted to verify the proposed technique. The outcome of the study revealed that the method was successful in detecting several types of structural damage for all three structures used in the study. However, the authors stated that the method might not work properly with ambient vibration data. It is also noted by the authors that the sensitivity of ARMA models of static response data to typical infrastructural damage is unproven and is to be investigated by the authors. Therefore, a generalization of the method cannot be made for other SDD applications.

To address the drawback of [130] for ambient vibration data, Gul and Catbas [131] suggested a modified nonparametric approach (a statistical pattern recognition methodology) capable of handling output-only signals recorded under ambient conditions. The authors proposed using RD method to compute the vibrations response per the ambient conditions before conducting the feature extraction process. Auto Regressive (AR) technique was then applied for time series modeling of the transformed signals. The features extracted from by AR were analyzed by a Mahalanobis distance-based outlier detector. Several experimental tests were carried out to examine the capability of the technique for SDD and localization. The laboratory experiments were conducted on a simply supported beam and a grid structure by random hand tapping excitation. As shown in Figure 13, all points that are below the threshold reveal that the numerically evaluated threshold value is consistent with the experiments. The experimental results demonstrated that the proposed technique can identify various damage cases by measuring dynamic response against unknown excitations. Yet, the authors noted that there are issues to be resolved before the method can be utilized in automated SHM applications. It was suggested that a robust method is required to improve the threshold value to accommodate more damage cases. A sensitivity analysis was also reported to be needed to investigate the effects of different parameters on the methodology. It is also emphasized that the effects of the environmental and operational variables need to be researched before the method is generalized for various other applications.

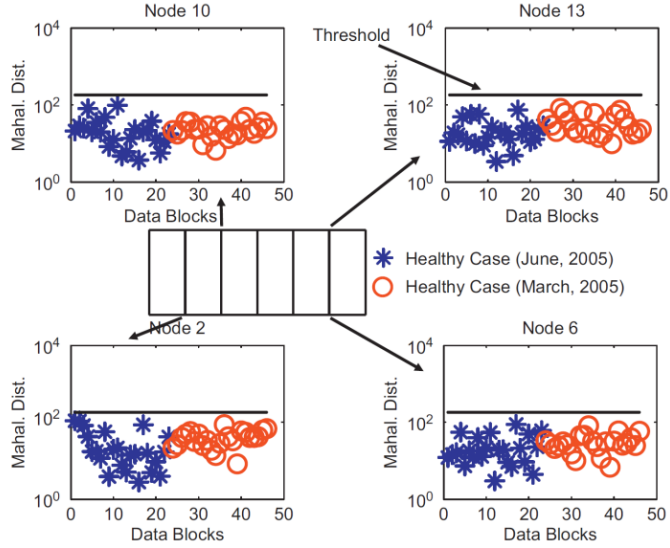

Figure 13 - Verification of the threshold value with experimental data in Gul and Catbas [131].

In a follow-up study, Gul and Catbas [132] introduced another SDD technique in which Auto Regressive eXogenous output (ARX) method was used for time series modeling and feature extraction. Rather than relying on a single model, an individual ARX model was used for each cluster of accelerometers on a grid structure (Figure 14). The damage features computed for each cluster were compared against those extracted under the undamaged structural condition to identify the presence and location of the structural damage. The method worked successfully for most of the cases for detecting and locating damage. This technique was tested experimentally on a laboratory grid structure and numerically on a simulated benchmark bridge structure (Z24 Bridge in Switzerland). The output-only data from the Z24 Bridge is used to successfully verify the method for various levels of pier settlements. The success of this approach with both laboratory and large-scale structure enables it to be generalized and used for other damage assessment studies for ambient vibrations.

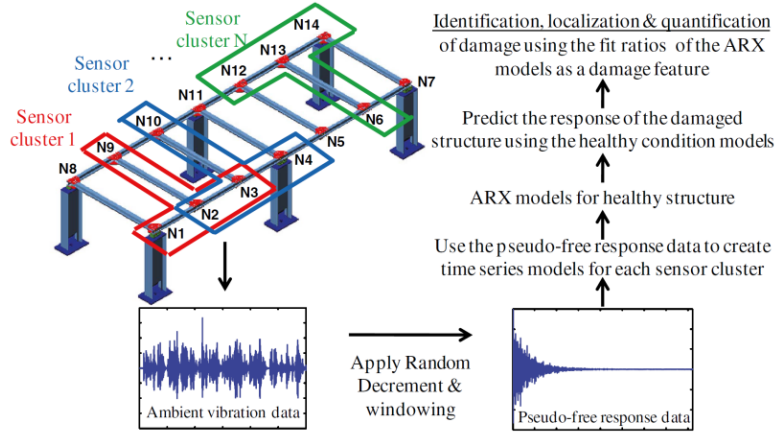

Figure 14 – Methodology used in Gul and Catbas [132].

Similarly, Ay and Wang [133] developed a SDD technique depending on ARMAX (Auto-Regressive Moving Average with eXogenous input) models fitted to the measured signals using an iterative self-fitting algorithm. These models were integrated with a multi-sensor data fusion technique to minimize measurement noise and therefore enhance the reliability of the proposed method. The ARMAX model is expressed as follows:

$$y(t) = \sum_{i=1}^{n_a} a_i y_{t-i} + \sum_{j=1}^{n_b} b_j u_{t-j-\theta} + e_t + \sum_{k=1}^{n_c} c_k e_{t-k} \qquad (6)$$

where $y(t)$ is the current output; $n_a$ is number of poles, $n_b$ is number of zeros, $n_c$ is estimate of disturbance using sigmoid network nonlinearity; $y_{t-i}$ is system output at time index $t$-$I$; $u_{t-j-\theta}$ is manipulated variable index $t$-$j$-$\theta$; $e_t$ is noise sequences;

$e_{t-k}$ is error term at time index $t-k$; $\theta$ is time delay, $a_i$ is auto-regression portion of model; $b_j$ is MA portion of model and $c_k$ is exogenous input portion of model.

A laboratory structure (Figure 15) with different damage cases (on connections) was used to test the damage detection performance. For damage detection an unlabeled signal was included in a stored pool of damage scenarios where the accuracy of damage detection is improved by multi-sensor data fusion. It was reported that the proposed technique is reliable for SDD; yet it has some shortcomings for practical applications. The difficulty of supervised learning format, small damage conditions (small cracks and notches) and noise contamination in time series data are listed as shortcomings of the proposed method. Therefore, a generalization of the method cannot be made for other SDD applications.

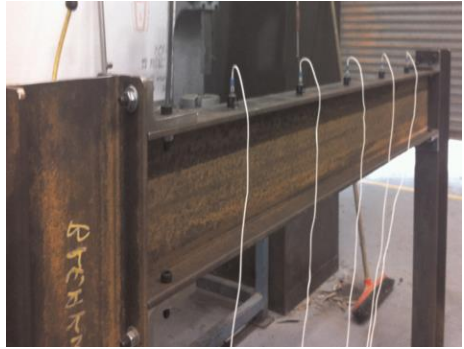

Figure 15 – The laboratory structure used in Ay and Wang [133].

Also, Yu and Zhu [134] reported another AR-based technique that employs fuzzy-c-means (FCM) clustering for AR parameters. The output of FCM was used to define six damage indices. The performance of these indices was verified using experimental data of a three story laboratory structures. It was reported that the proposed method was successful for nonlinear damage detection; and environmental variability has the potential to affect the residual errors standard deviation ratios. On another note by the authors, it was suggested that the proposed procedure is verified with additional experimental studies possibly by large-scale structures.

Goi and Kim [135] proposed a nonparametric method that utilizes a Multivariate-AR (MAR) approach to model the ambient vibration response of bridges. Damage detection was conducted through statistical hypothesis testing per the baseline condition of the actual steel truss bridge with truss members that were artificially damaged (Figure 16). The damage indicator assessed the stochastic distance of the principal components between groups of references with the Mahalanobis Distance (MD) method. On the MAR model, Principle Component Analysis (PCA) was utilized to extract damage features. As a next step, statistical hypothesis testing was utilized according to a probability distribution of the damage indicators for SDD. With the experimental data recorded on the bridge, the damage indicator successfully detected three damage cases but the performance was observed to depend on the AR order. Comparisons with different univariate AR approaches indicated that the proposed technique is efficient for SDD and localization. Based on the successful performance on an existing large civil structure, the approach has the potential to be used efficiently on other applications.

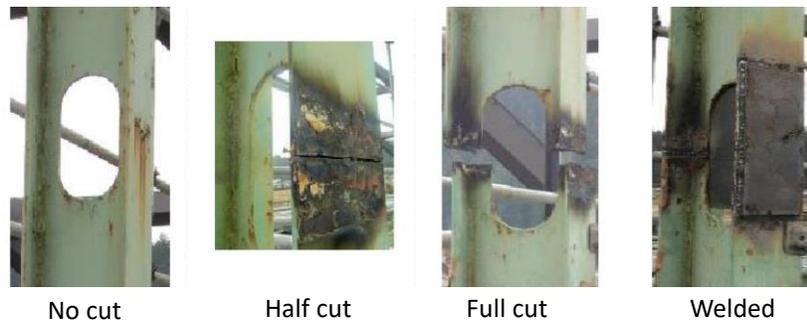

Figure 16 – Damage cases on the flanges of vertical truss member at the bridge midspan in Goi and Kim [135].

More recently, Datteo et al. [136] applied a nonparametric technique for SHM and SDD of the Meazza Stadium (Italy) from August 2015 to April 2016. The authors utilized AR models to model the accelerations response of the stadium in terms of 40 parameters. PCA was then utilized to decrease the dimensionality of the extracted feature vector and describe the signal characteristics in terms of fewer independent variables. The first three principle components were used as indicators to changes in weather, environmental and structural conditions. It is reported that there is good correlation between humidity, temperature and the principal components under consideration. The first principal component was affected more per changes in weather conditions, whereas the other two components were more influenced by operational conditions. The authors highlighted that it is needed to filter-out and segregate the effects of deterministic weather conditions and operational circumstances on the principal component scores for the SDD task to be more accurate and effective.

On a comparative study on experimental assessment of several statistical time series methods, Kopsaftopoulos and Fassois [137] investigated the efficiency of vibration based statistical time series methods for SDD via a lightweight aluminum truss structure tested in the laboratory (Figure 17). The study involved a comparison of the major non-parametric (Power Spectral Density, Cross Spectral Density, FRF magnitude based methods) and parametric (model parameter, residual variance, residual likelihood, residual uncorraletedness based) techniques under various excitations and multiple damage cases based on scalar versions of the techniques. The results of the study confirm the high potential and effectiveness of statistical time series methods for SDD. It is important to note that the parametric methods have shown almost perfect performance with no false alarms, missed damage, and damage misclassifications. Parametric time series methods were found to be more detailed in procedure and required more expertise as opposed to their primarily simpler non-parametric counterparts. As an example, the FRF magnitude based SDD results are shown in Figure 17 indicating that the test statistics is not exceeding the threshold value for the undamaged state while exceeding for the damaged states. The authors emphasized the fact that the generalization of the methods requires the use of corresponding vector models and multivariate statistical decision making. In addition, the methods need to be verified on large-scale structures subjected to uncertainties and different environmental conditions.

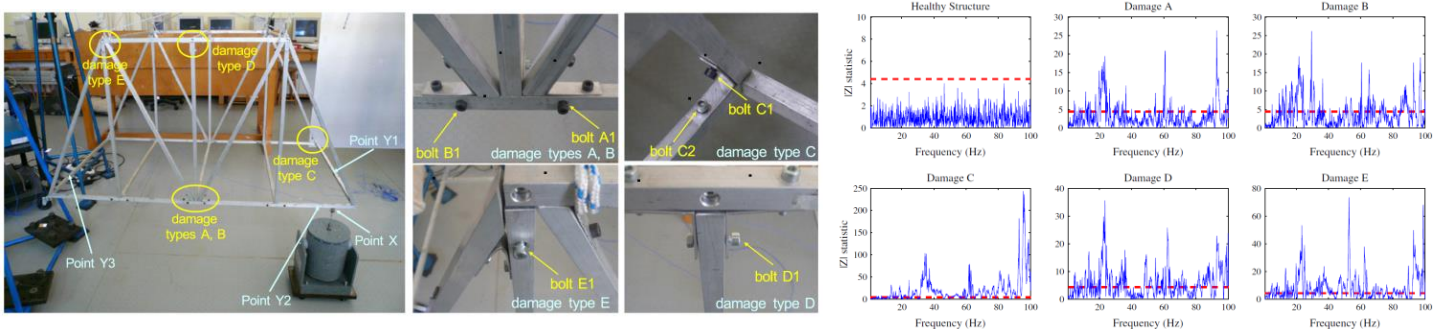

Figure 17 – The laboratory truss structure and the damage locations (left); the FRF magnitude based SDD results (right) in Kopsaftopoulos and Fassois [137].

A novel nonparametric and unsupervised SDD methodology based on a GA is proposed by Silva et al. [138]. The SDD technique is per genetic algorithm for decision boundary analysis (GADBA) which is designed to operate under operational and environmental variability. Searching for an optimal number of clusters in the feature space, the GA based clustering approach is based on a novel Concentric Hypersphere (CH) algorithm for regularizing the number of clusters and mitigating the cluster redundancy. Two real-world datasets, the Z-24 bridge (Switzerland) and the Tamar Bridge (UK), were used to compare the performance of the GADBA method to the two state-of-the-art parametric cluster-based approaches; the Gaussian mixture models (GMM) and the Mahalanobis squared distance (MSD) (Figure 18). The results revealed that the proposed GADBA method has a better classification performance than the others since in GADBA method the genetically guided characteristic increases the chance to get a solution close to the global optimal. For GMM however, Gaussian distributions and the Expectation-maximization (EM) algorithm converge toward a local optimum. As such, GADBA methodology is a better fit for real-world deployment.

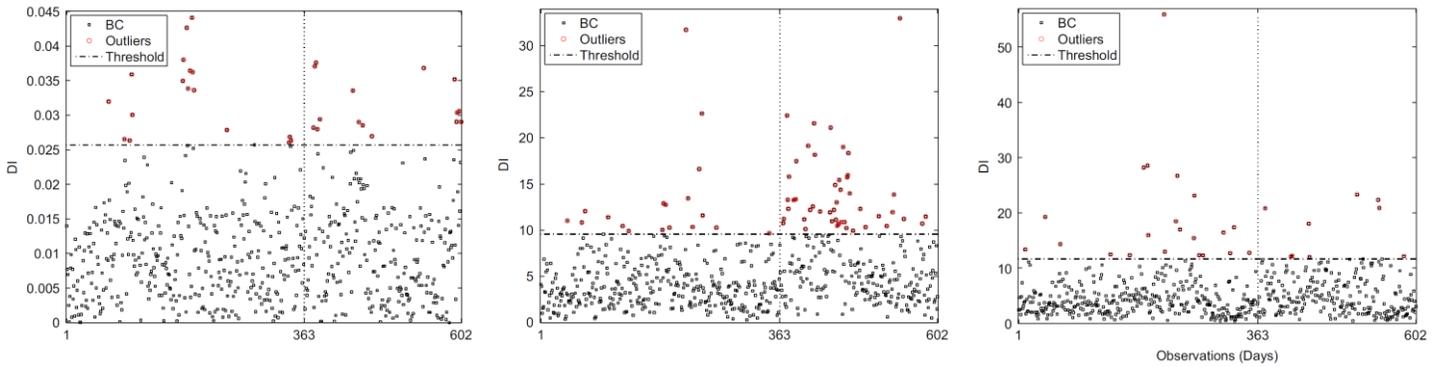

Figure 18 – Tamar Bridge Damage indicators along with a threshold based on a cut-off value of 95% over the training data: (a) GADBA, (b) GMM-and (c) MSD-based approaches, in Silva et al. [138].

## 4. Vibration-based structural damage detection methods based on Machine-Learning

Based on the extracted information from the measured signals, vibration-based SDD techniques have been categorized into nonparametric and parametric methods which have been discussed in detail in Section 3. Based on this categorization, the methods reviewed in Section 3 were the automatic and unsupervised methods which are not based on machine learning (ML). In this section, supervised vibration-based SDD methods based on ML algorithms will be reviewed.

It is crucial to note here that the model-driven methods require manual labor from the experts who have a deep knowledge on the physical, mechanical, electronic, data flow, or other appropriate details of the complex system. So, such experienced experts are usually both scarce and expensive resources, and on top of that they are always prone to human error. The most crucial drawback is that human-centric modeling usually takes considerable time. It is inherently a trial-and-error approach, rooted in the old scientific method of theory-based hypothesis formation and experiment-based testing. Finding a suitable model and refining it until it produces the desired results is often a lengthy and costly process. Finally, the model-based approaches are usually static, cannot adapt to a change or new occurrences of system (or problem) parameters if they were not the part of the "model" that was designed initially. When such a change occurs, a "re-modelling" is usually needed, which means an update of the old model or sometimes, a new modelling from scratch - again by the experts/professionals.

Data-driven methods are based on a ML paradigm and as the name implies, they are "data-hungry". Before a decade or so collecting "useful" data (labeled and organized) for training was not easy and cheap; however, with the current technology level the World has reached today, this is no longer a bottleneck and the advances in AI and DL domains yield numerous data-driven methods trained over large-scale datasets. Once trained, the data-driven method can be used as a "black-box" to provide a generic and dynamic solution which is optimized by the underlying training method for the problem at hand. When the need arises, the "black-box" can incrementally be updated according to the changes in the system parameters. In the recent literature numerous methods have been proposed which can achieve expert-level or even beyond performance levels for several anomaly- or fault-detection problems. They are now quick and cheap to create and on top of this a data-driven method trained for a particular problem can even be used for another, perhaps totally independent problem via the paradigm called "transfer learning".

As discussed earlier, ML enables the systems to automatically learn and improve from experience without being explicitly programmed. This happens while the ML algorithm allows the computer to learn the knowledge required for conducting a specific task by analyzing a sufficient number of relevant data. Traditional ML algorithms require the data representation in terms of a fixed number of features. Therefore, a "feature extraction" step is needed which is simply preprocessing the data to extract the certain attributes which represent the most characteristic information. Together with the data labels, the processed data samples expressed in extracted feature terms are utilized to train the ML system based on a specific training method in order to learn how these features correlate with different data patterns.

ML algorithms have become very popular and broadly utilized in numerous vibration-based SDD methods. Like the non-ML-based methods, ML-based SDD methods are categorized into parametric and nonparametric methods. A large portion of parametric and nonparametric ML-based SDD systems perform the two common tasks: feature extraction and training.

Then the trained ML system is utilized to identify the presence and location of structural damage by performing classification. Based on this overview, following two subsections present the review of recent applications of ML algorithms in parametric and nonparametric vibration-based SDD techniques. It should be emphasized that a particular consideration is drawn onto feature extraction and feature classification techniques which are utilized to perform the SDD operation.

**4.1 ML methods for parametric vibration-based SDD**

Several parametric ML-based SDD methods are proposed in which the feature extraction process is carried out by simply identifying certain modal parameters from the structural systems using input-output or output-only modal identification techniques. A well-trained ML classifier is then used to process the extracted modal parameters to assess the structural integrity. The most commonly used parametric ML-based approaches are those that rely on modal characteristics such as natural frequencies and mode shapes as extracted features along with the feed-forward, fully-connected, multi-layer Artificial Neural Networks (ANNs) or the so-called Multi-layer Perceptrons (MLPs) as classifiers.

An integrated method of Spatial Fourier Analysis and ANNs was proposed by Pawar et al. [139]. The authors developed a vibration-based method for SDD and damage localization in fixed-fixed beams using a damage index in the form of a vector of Fourier coefficients. Spatial Fourier analysis was conducted to identify the mode shapes (the extracted features) based on free acceleration response of the beam. An MLP was trained and used for processing the extracted mode shapes (first three bending modes only, as shown in Figure 19) to find out the damage location. The method was verified analytically using an FE model of the fixed-fixed beam. It is reported that the parameters like number of neurons in the hidden layers, the structure of hidden layers and input–output data scaling are achieved by varying each parameter within certain limits. With that, the network was trained to achieve the most optimum condition leading to minimum error. Meanwhile, per the rest of the numerical findings, it is reported that a linear activation function is formed and utilized between output and hidden layers. One hidden layer is found to be the most efficient; which resulted in the number of neurons for the hidden layers to be chosen as the following: 100 for 63 inputs; 75 for 45 inputs and 50 for 33 inputs. While the Fourier coefficients are found to be sensitive to both damage size and location; higher harmonics are found very sensitive to damage location which means the Fourier coefficients in the spatial domain can be considered as damage indicators. As such, the ANN results revealed that this method can detect, locate and quantify small damage levels through noisy data. Even though the method is recommended to be used in online SDD applications, it would be safer if the method is evaluated and verified on a real structure before a generalization can be made.

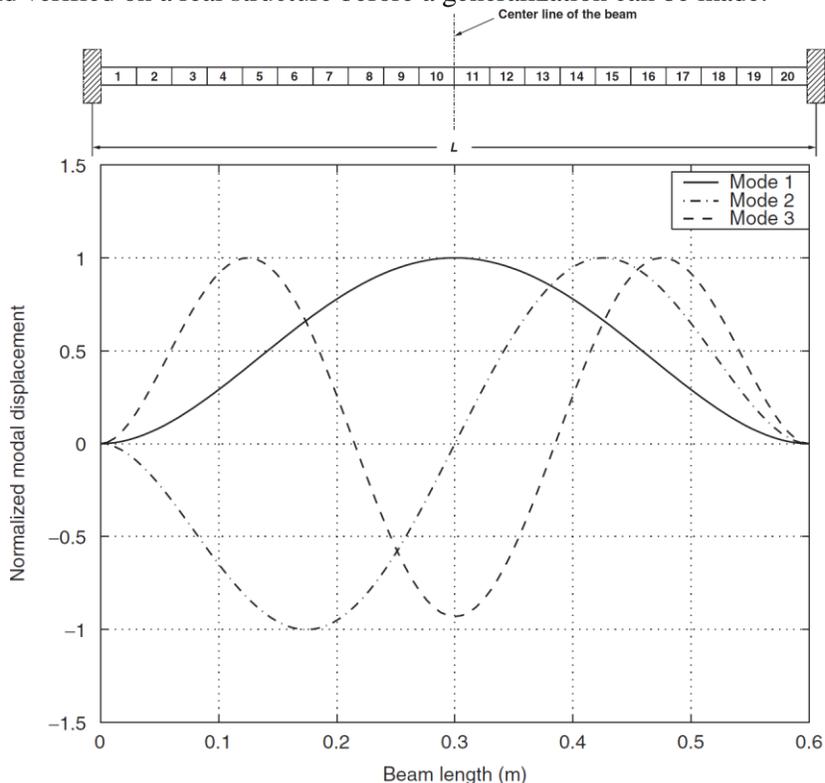

Figure 19 – Fixed-fixed beam and the first three bending modes used in Pawar et al. [139].

Mehrjoo et al. [140] proposed an ANN based SDD technique on a simple truss bridge and the Louisville Bridge truss in which modal characteristics were extracted from acceleration response and utilized as damage sensitive features. An MLP with a single hidden layer was used for damage identification and localization. While the value in each neuron indicates the damage percentage (DP) of the node, 273 training patterns for the simple truss and 729 training patterns for Louisville Bridge truss are arranged randomly before training. The first four bending modes were used for the training process of the simple truss and the first five modes were used for the training process of the Louisville Bridge. Damage was introduced simply by stiffness reductions on certain truss members. The complete training process of networks needed 75k epochs using the standard back-propagation algorithm. The learning curve for the Louisville Bridge truss is shown in Figure 20. Based on this figure, the RMS difference vector for training and testing patterns decreases when the number of epochs increase; however, RMS difference vector for testing patterns starts to increase very slightly after 75k epochs. The reason for this is the network becomes over-trained and tends to lose its generality when the number of epochs exceeds a specific limit (optimal epoch). For the most efficient configuration, it was reported that the number of output layer neurons was 5; the number of input layer neurons was 32; the number of hidden layer neurons was 50; learning coefficient rate was 0.5; learning coefficient was 0.95; and the momentum rate was 0.3. When the technique was tested by numerical simulations conducted on the simple truss structure as well as the FE model of Louisville Bridge truss (Figure 20), the damage and its severity were identified with good precision. Even though the authors recommended the use of this approach in on-line and real-time SDD applications, it would be safer to test and validate the method on a real structure first.

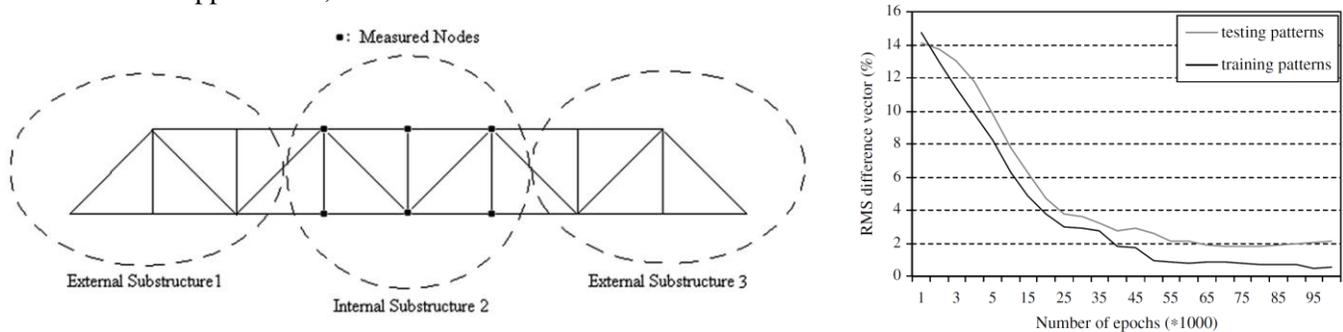

Figure 20 – Louisville Bridge truss (left) and the variations of RMS difference vector of outputs for learning and testing patterns (right) Mehrjoo et al. [140].

On a study on the complexity of ANNs for SDD, Yuen and Lam [141] used the modal parameters of a simple five story structure with a numerical model along with a MLP (with 5 input neurons and 5 output neurons) for detecting damage. The motivation of the authors was the fact that even though the number of hidden layers and hidden neurons in each hidden layer has crucial effects on the training, the Bayesian probabilistic method to select the ANN model class with suitable complexity is often overlooked. As such, a two-phase damage detection method and a Bayesian ANN design method are utilized. The number of input–target pairs used for the training is reported to be 32. For simplification, ANNs with one hidden layer is used first, by which the authors reduced the ANN design process to the number of hidden neurons. Different numbers for hidden neurons were also studied and it was concluded that the ANN structure with 6 hidden neurons is the optimum choice. The selection of the ANN architecture was carried out systematically using a Bayesian ANN design approach. The damage scenarios were defined by applying 20% to 80% inter-story stiffness reduction at various locations of the five story model. The numerical results confirmed the capability of the proposed technique to identify both single and multiple damage scenarios. For example, for a single damage case, 19.7% of ANN output is achieved for a corresponding simulated damage extent of 20%. The illustrative example of this method can be used to design an ANN with any number of hidden layers and detect damage for multiple-damage cases; however, the methodology needs to be tested and verified on a real-structure before it is generalized.

An extended Bayesian ANN approach was introduced by Ng [142] to determine the structure of the ANN used for SDD and damage localization. Numerical data from a simulated benchmark problem was used for training and testing the ANN. The feature vectors used in this study were simply the modal characteristics of the benchmark model under several structural damage states. The number of damage states was reported to be 113. In an extended Bayesian ANN algorithm noted by authors, ANNs with the tangent sigmoid transfer function and 17 neurons in the hidden layer resulted in the highest logarithm of evidence, which is the main reason why they are used in the ANN architecture for SDD in [142]. It is noted that there is a small change in the modal parameters when damage is present (Figure 21). The identified modal frequencies and mass normalized mode-shapes are utilized to compute the damage-induced changes on modal parameters, and are then used as input items to train ANNx and ANNy to identify the horizontal stiffness reductions in two orthogonal directions for the damage cases. As such, the proposed method resulted in satisfactory performance for SDD on the benchmark structure. As an example result, when the horizontal stiffness reduction at the first story is 11.31%, the damage detection tool resulted in a horizontal stiffness reduction of 10.96% at the same story. The performance of the SDD algorithm was found satisfactory for all six damage cases considered in the benchmark study. Since the data used in this study was from a large-scale structure and since it is validated for all damage cases of the benchmark study, it can be generalized to be used for other applications.

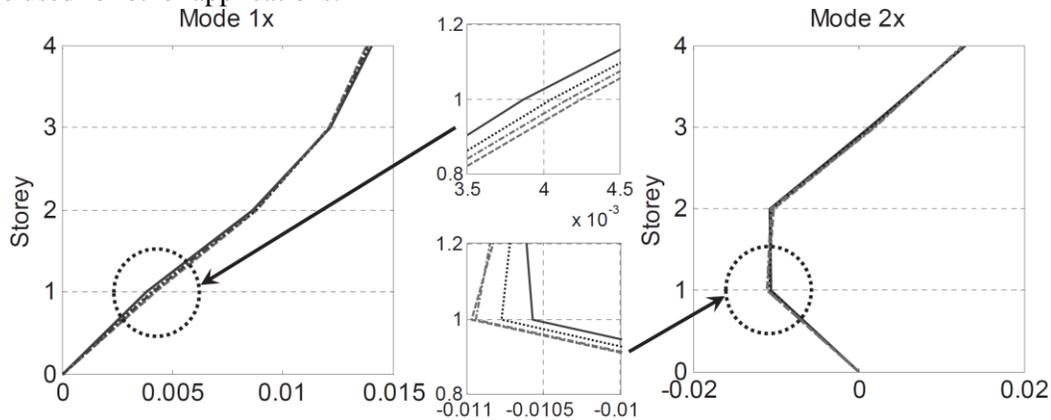

Figure 21 - Identified mass-normalized mode-shapes in the x-direction and zoom-in on the 1$^{st}$ story for four damage cases in Ng [142].

Instead of using a single ANN to handle the feature classification process, Gonzalez and Zapico [143] suggested the use of multiple ANN classifiers to improve the damage detection performance. They proposed a ML-based parametric technique for SDD that relies on an ensemble of two ANNs for seismic damage identification intended for a steel moment-frame representing a five-story office building (Figure 22). After the calibration of the initial stiffness and mass of the structure, the following stage is the identification of the final stiffness and mass after a severe earthquake. The inputs to the ANNs (i.e. damage features) were the modal parameters computed for the first three modes, while the outputs were the modal mass and the modal stiffness of structural members. The results of a FE model are used for training of the ANNs. The evaluation of the algorithm reveals that the damage predictions for unseen random data are in good agreement with the target values. On another note, the statistical analysis showed that the method is sensitive to modal parameter errors. Further investigation on modal parameter errors revealed that the coefficient of variation of the modal errors should be less than 0.1% for natural frequencies and 0.02% for mode shapes to obtain absolute values of the damage prediction errors up to 0.05 with a 95% confidence. While this method was able to predict changes in mass and stiffness directly from the extracted modal parameters, its performance needs to be verified on a real building before it can be generalized.

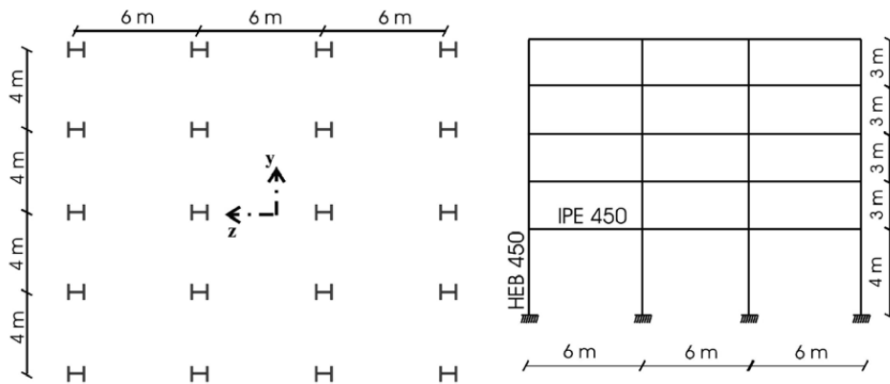

Figure 22 – Five story building used for the simulations in Gonzalez and Zapico [143].

Another multi-ANN based approach was introduced by Bakhary et al. [144] who developed a two-stage ANN classifier (Figure 23) in an attempt to minimize the computational work for SHM and SDD of a large structure. A two-span reinforced-concrete slab which is 6.4 m long, 0.8 m wide with 10 cm thickness (Figure 23) is assumed for the numerical study. Four damage scenarios involving damage in single and multiple substructures are simulated to test the ANN performance by measuring 33 points on the centerline along the 6.4 m length. The modal analysis is run per the FE model and the first three frequencies are used for both deterministic and statistical methods. The idea of this approach is to divide the infrastructure into several parts and then train an ANN to compute the local frequencies of the substructure in terms of the modal frequencies. The authors calculated the probability of damage existence (PDE) by comparing the probability distribution of the damaged and undamaged models. The uncertainties are considered in both the training and testing data and assumed to have zero means; while the errors are assumed as Gaussian distribution and varying around the true values randomly. The output of the ANN is processed by a second ANN to localize and quantify structural damage. Finally, the results of the FE model are compared to the laboratory tested reinforced concrete slab for the SDD procedure. The results showed that the proposed method is successful at detecting small damage with a higher level of confidence. On another note the authors verify the suggestion by Trendafilova et al. [145] on the fact that SDD via the substructuring technique needs to be approached in terms of probability of damage rather than deterministic determination of damage levels. Since the approach is validated numerically and experimentally on a relatively large structure, the approach has the potential to be used on other applications.

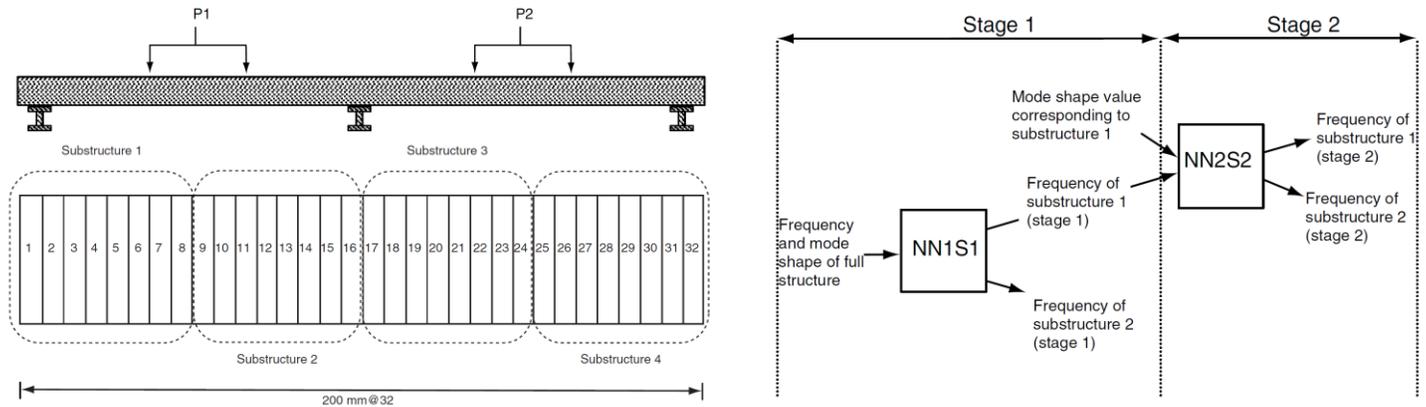

Figure 23 – Slab elevation and plan (left) and two-stage ANN model (right) used by Bakhary et al. [144]

Lee and Yun [146] also verified their ANN-based damage detection approach both analytically and experimentally. A single ANN was trained to identify structural damage according to the mode shape changes with respect to undamaged condition. The ANN was trained based on an FE model of Hannam Grand Bridge in Seoul, South Korea. Then, field measurements collected from the same bridge under ambient vibration were used to test the performance of the ANN under three damage cases. The network performance with strain signals was found to be relatively more successful as opposed to using acceleration response. Furthermore, the misclassification rate was observed to be decreasing with increasing damage severity. However, higher misclassification rate values were observed in higher damage severity cases (which is tied by the authors to the excessive FRF shape changes). The results of damage localization per probabilistic neural networks (PNN), sequential PNNs and back-propagation neural networks (BPNN) are shown in Figure 24. In conclusion, PNN using sequential estimation scheme was successful at detecting multiple damages. While the conventional BPNN is an alternative to detect multiple damages, its performance for single damage case resulted in false alarms. As such, PNN method is found to be better fit for complex structures than the conventional ANNs. While the approach in [146] is very promising, the number of false alarms discussed in the text is preventing the method to be generalized for all practical SDD applications.

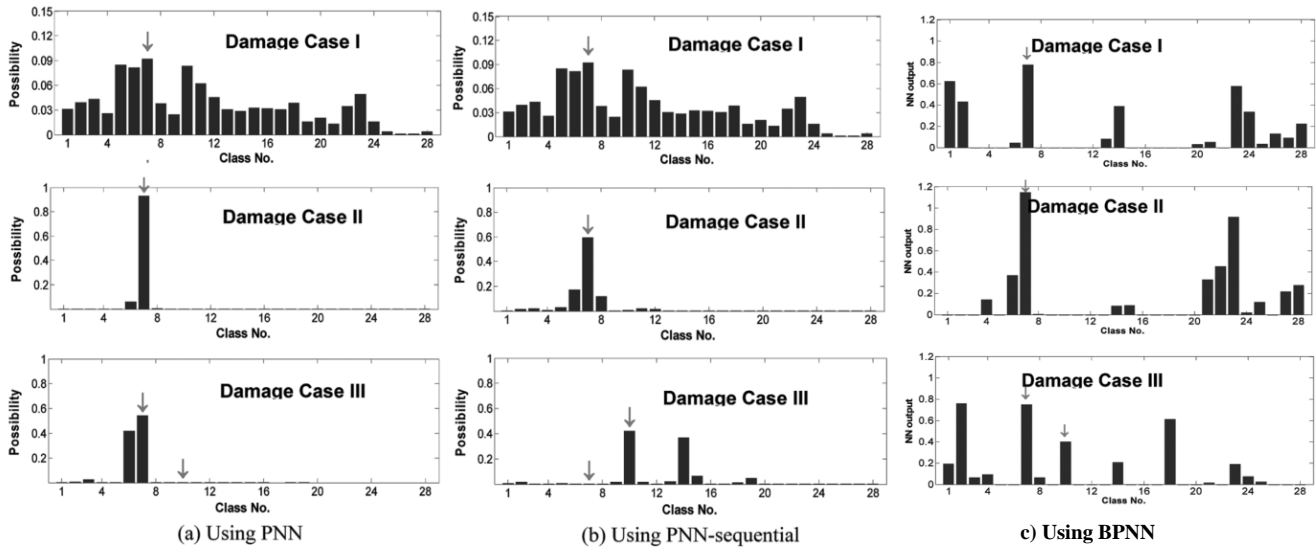

Figure 24 – Damage localization plots by Lee and Yun [146].

In another experimental work, Hakim et al. [147] used an I-beam with fixed-fixed support conditions (Figure 25) to test an ANN-based SDD. Damage was introduced by grinding slots on the beam flange at different locations and depths. Accelerations along the beam were measured under a white noise excitation applied using an electrodynamic shaker. A modal estimation technique was used to determine the natural frequencies for the first five modes corresponding to the undamaged and damaged scenarios. The data was used to train an ensemble of five ANNs to correlate the location of the damage and the severity of the damage to the changes in the natural frequencies. Based on the training and validation stages, the most efficient system was observed to be the case where two hidden layers are used (seven neurons in the first hidden layer and six neurons in the second hidden layer) as a result of 16036 iterations. On another note, the individual ANN tests for double damage scenarios were reported to result in inaccurate estimations for locations of damage in very-light damage scenarios (since the damage was not generating recognizable change in the modal curvature). However, the damage severity predictions were successful for all damage cases. In addition, it was reported that the ensemble of ANN predictions were accurate for very-light damage scenarios, and they were observed to be in high accuracy even when the trained datasets were noisy; the damage detection and quantification was successful even for double damage scenarios (Figure 25). Based on the presented results in the paper, the methodology can be generalized for other applications only after verification on a large-scale structure.

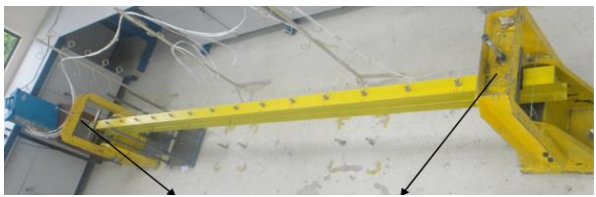

Fixed supports at both ends

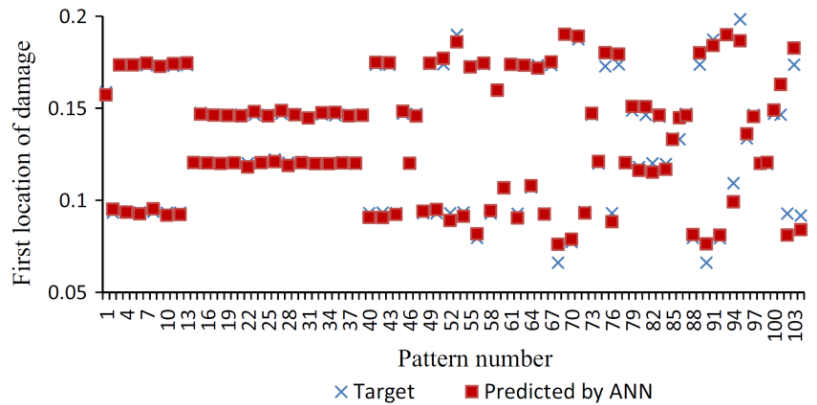

Figure 25 – Experimental beam (left) and SDD predictions by ANN (right) in Hakim et al. [147]

On another work utilizing ANNs along with genetic algorithms, Betti et al. [148] used field measurements obtained from an experimental benchmark SHM problem of [129]. The damage level on the structure is increased by progressive cutting of column flanges (Figure 26). The modal characteristics were obtained based on the ambient vibrations utilizing an output-only modal identification method. The employed ANN was a feed-forward back-propagation (FFBP) network structure with two hidden layers and it was trained to carry out the damage classification process. The input layer contained the input parameters with four neurons, each processing one of the frequency-dependent indexes. The output layer processed a linear output combination for the second and third layer of neurons. With that, the network output was able to determine a resonant condition. The authors used sigmoid activation functions to characterize the hidden layers while the performance of networks per hidden layers with five, ten or fifteen neurons was monitored. The data required for training the ANN was generated using an FE model of the benchmark (Figure 26). Two fitness functions were employed and the results of the experimental results were compared with the results of the optimization algorithm to ensure its capability to match the actual damage using the first four modes. The authors used a genetic algorithm (GA) to update the FE model to make it more representative of the actual structure. The authors concluded that a combined ANN/GA method for SDD is a powerful technique; they also report that the modal assurance criterion was successful to form a fitness function for GA. Since the methodology worked successfully per the real data of benchmark structure, it should be able to be used effectively on another structure.

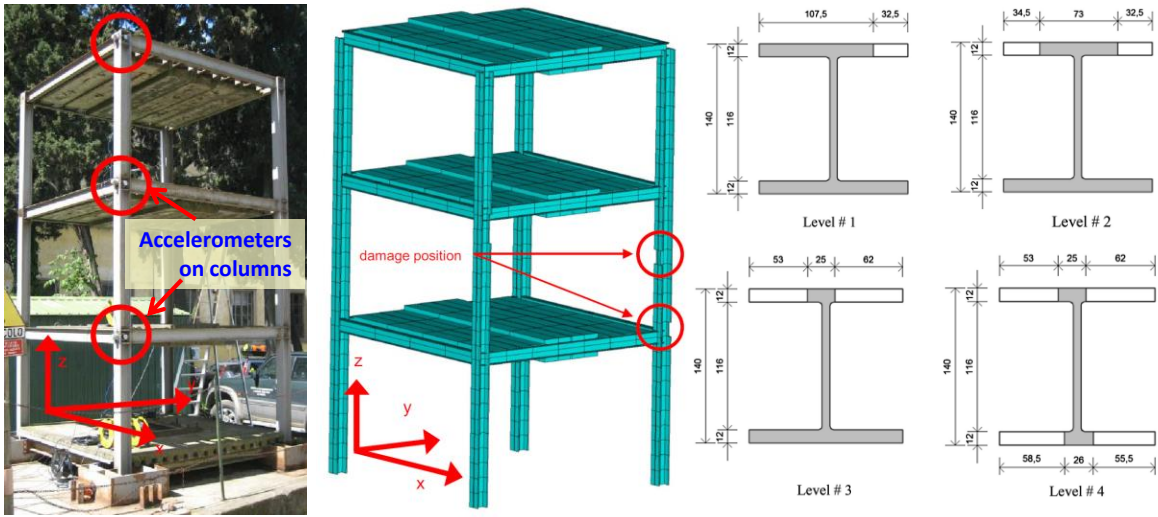

Figure 26 – Benchmark structure (left); FE model (middle); and induced damage on column cross-sections (right) used in Betti et al. [148].

In some studies, researchers have combined modal identification with other techniques for running the feature extraction process. In this context, Rucka and Wilde [149] proposed an ML-based damage detection system that uses modal testing along with wavelet transformation for extracting damage-sensitive features. The method requires estimating the mode shapes from the ambient response. The computed modes are then processed using Continuous Wavelet Transform (CWT). An MLP with two hidden layers (100 neurons and 20 neurons in the first and second hidden layers, respectively) was trained for SDD and damage localization using the CWT method. The performance of this method was evaluated experimentally using a small beam, plate, and a shell structure. Based on the presented results, the defect presence identification by the CWT was possible only for 5 of 12 input patterns. The Neuro-Wavelet (NW) system prediction was completely successful for the 10 patterns and reasonably successful for 2 patterns. The NW system was successful to identify the damage presence for all considered damage states with an average error of 3.17% (Figure 27). The authors concluded that the wavelet damage detection method would be suitable for detecting relatively small damages (with damage area being 0.2% of the total shell area). It was also reported that, in more complex structures, the sensitivity of higher modes to small defects might be high (the shell's highest considered mode appeared to be the best for damage detection). Considering the size of the tested structures and based on the presented results in the paper, the methodology can be generalized for other applications only after verification on a large scale structure.

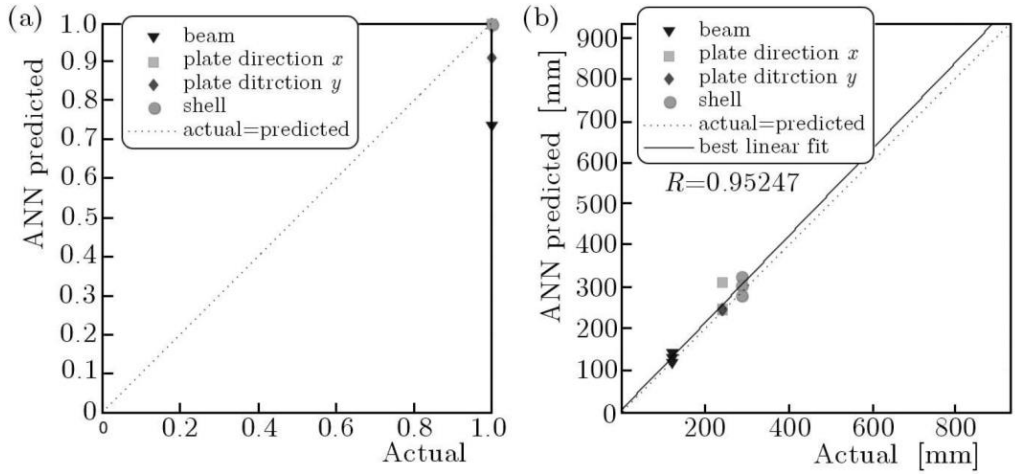

Figure 27 - Results of the Neuro-Wavelet system for the experimental beam, plate and shell mode shapes; (a) crack presence, (b) crack localization per Rucka and Wilde [149]

On another parametric ML-based study, Lam and Ng [150] presented a method where the modal parameters and Ritz vectors were used simultaneously as extracted features. A single MLP optimized through a Bayesian approach was then used to classify these features considering both the optimal number of hidden neurons and the activation function in the hidden layer. An analytical benchmark model [129] was used to evaluate the technique for SDD and damage localization. The proposed algorithm ensures that the trained ANN shows good performance while being simple at the same time (avoiding data over-fitting and fluctuation in output). While it is reported that "tansig" is a better activation function in the hidden layer than "satlin" in terms of SDD, it is noted that ANN design and SDD method are verified using the first five cases of the benchmark study. Four MLPs are used; two of which are trained by damage-induced modal parameter changes and two of which are trained by damage-induced Ritz vector changes. Figure 28 shows the logarithm of the evidence, $log(e(a, n))$ for $a = 1$, for different numbers of hidden neurons. It is noted that as the number of hidden neurons increases, the value of the evidence increases until $n =17$. There is indeed a drop after this point, which makes $n = 17$ is the 'optimal' number of hidden neurons for this case. As a result, the stiffness reductions were successfully identified by trained MLPs. It is also reported that the performance of MLPs trained by modal parameters resulted in better performance than MLPs trained by Ritz vectors. Since the methodology worked successfully per the benchmark model, it should be able to be used effectively on another structure.

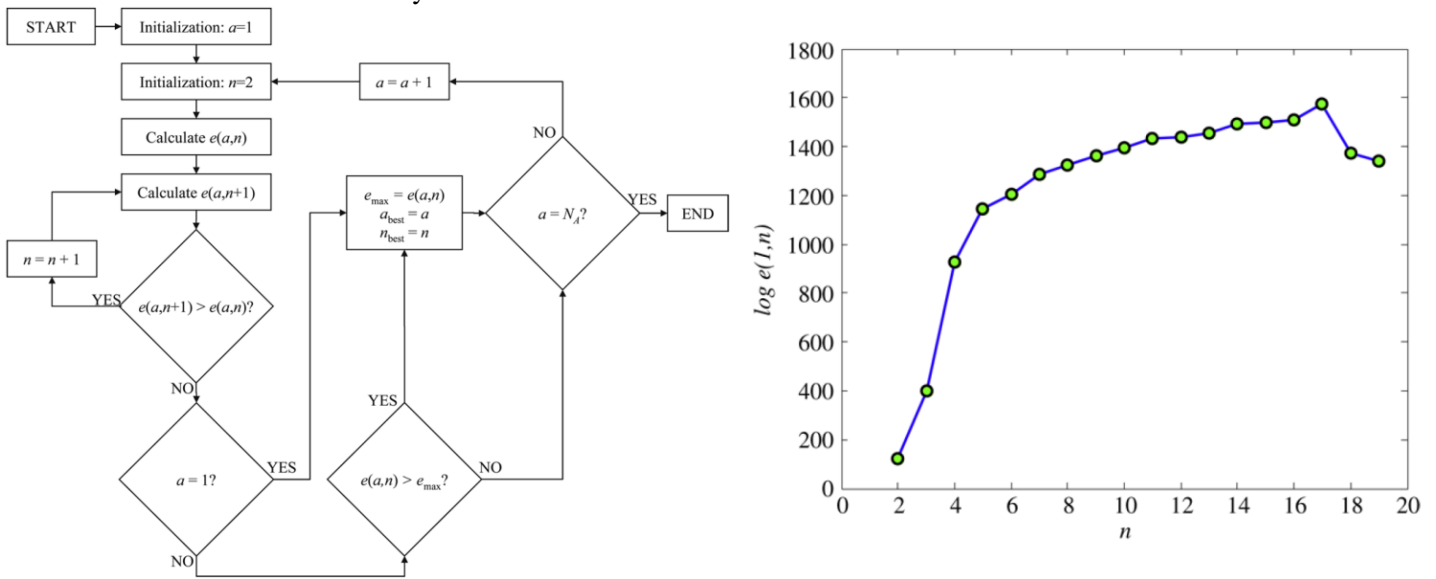

Figure 28 - ANN design algorithm for the 'optimal' number of hidden neurons and the 'best' activation function (left); iteration history of selection of the optimal number of hidden neurons with "tansig" activation function $(a = 1)$, in Lam and Ng [150].

In addition to MLPs, few parametric ML-based methods have employed other types of ANNs for carrying out the feature classification process. For example, Jiang et al. [151] used probabilistic neural networks (PNN) for classifying the modal parameters extracted from the vibration response. A new eigen-level data fusion model for SDD is investigated combining the advantages of rough set processing, PNNs and data fusion. Two numerical examples revealed that the proposed model is effective. The fusion model resulted in successful SDD and noise-resistant capabilities while reducing the memory requirements. It is observed that the combined use of rough set processing, probabilistic neural networks and data fusion has a great potential in SDD; yet, it should be emphasized that the methodology has been verified on two numerical examples only. It is recommended that the method is verified on a real structure so that it can be generalized for further applications.

Another parametric ML-based study is by Lee and Lam [152] where the authors introduced an ANN model which is basically integrating the general regression neural network model (GRNN) and the fuzzy ART (FA) model resulting in "GRNNFA". While the model is fast and stable for network training and incremental growth of network structure, it also removes the noise the training samples. The modal parameter changes due to structural damage are used as patterns for GRNNFA model which was trained to learn for effective SDD. The methodology is found successful even in the presence of noise; however, future work was recommended to verify and validate the model on real structures so that it can be used for online and real-time applications.

In a study by Jiang et al. [153], the authors presented the ANN model with the motivation that the combination of fuzzy neural network and data fusion is more effective than any other SDD method. The proposed two-stage SDD methodology runs as the following. A primary assessment on FNN based SDD is conducted and then the final refined assessment is conducted utilizing data fusion and FNN models. The fusion decision-making model is the key for effective and feasible SDD task which resulted in increase in accuracy and decrease in uncertainty. Even though the results are very promising, it is worth mentioning here that the authors achieved them over numerical simulations. Therefore, the performance of the methodology needs to be verified on a real structure so that it can be generalized for other applications.

Wen et al. [154] presented a SDD study on unsupervised fuzzy neural network (UFN). The damage location per reduced stiffness is represented by the damage localization feature (DLF). In the presence of damage, a DLF is calculated and the damage location is identified via the UFN. The proposed model is validated by using a five-story structure subjected to single and multiple damage cases. The damage cases are simulated by reducing the stiffness of the first, second or third floor. Based on the measured DLF information, the UFN issues the damage location (for single and multiple damage cases) which is verified numerically for a five-story structure. Even when noisy and incomplete modal data are introduced, UFN was successful in SDD tasks. However, it is recommended that the method is validated on a real structure before a generalization can be made.

Other classifiers besides ANNs have been also utilized for parametric damage detection. For instance, Meruane [155] utilized an ML technique entitled Online Sequential Extreme Learning Machine (OSELM) to classify extracted modal parameters. The method was tested experimentally under several damage cases introduced on a rectangular beam (Figure 29) and a mass-spring laboratory structure. An MLP with one hidden layer was trained with data per a numerical model which was later run with real data for two real damage scenarios. The authors compared the findings against a GA based updated model. The algorithm was effective for SDD per transmissibility approach and its anti-resonant values, for both structures (Figure 29). It was also reported that the proposed technique successfully predicted the minimum damage levels of 30% for the beam structure and 20% for the mass-spring structure. It was observed that the levels of damage were not accurately predicted when the MLP is used. The author suggested that the MLP is a better match for rapid SDD while the GA is a better match for cases where a precision in SDD is required.

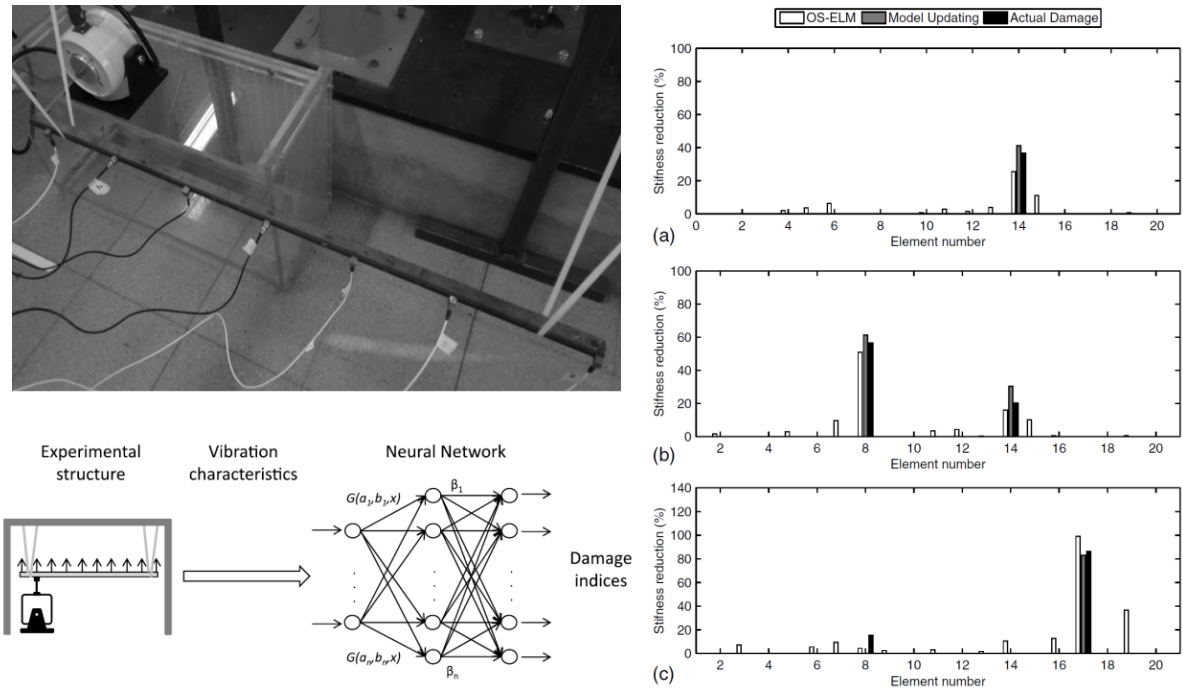

Figure 29 – The experimental setup and the methodology (left); SDD in the beam using OSELM and model updating for three damage cases in Meruane [155].

Another example for parametric ML-based techniques involved pattern recognition of structural health based on learning algorithms and symbolic data. In this work, Cury and Cremona [156] conducted a comparative study to evaluate feed-forward ANNs, SVMs, and Bayesian Decision Trees (BDT). The authors utilized Symbolic Data Analysis (SDA) to manipulate ambient condition acceleration recordings and modal properties of a steel railway bridge in France (Figure 30) which was monitored to characterize and quantify the effect of a structural strengthening. This strengthening was about installing special bearings (Figure 30) to move the first natural frequency away from the excitation frequency of trains passing by, to avoid resonance. Excitation tests were performed before, during and after the strengthening operation. Both raw acceleration data and modal information was used for feature extraction. It is reported that the classical analyses were not successful for SDD. When the training data set comprises 30% of tests, it was reported that only the SDA-SVM method reached 100% accurate classification performance. The SDA-BDT method did not reach more than 90% detection probability. The authors concluded that noisy data makes it difficult to achieve a satisfactory classification performance. Ten thousand simulations were run and SDA-NN and SDA-SVM methods were found to be successful in the classification of 100% of the cases. When it comes to the mode shapes, the SDA-NN method and the SDA-SVM method was reported to perform better than the SDA-BDT (which was attributed to the fact that the SDA-BDT is not an ML-based method). Since the methodology is verified on a large-scale structure, a generalization can be made on its use for other applications.

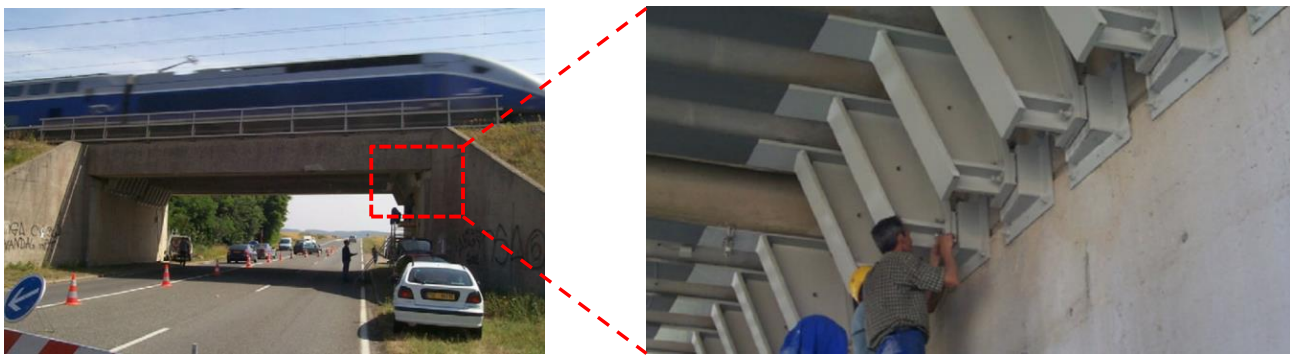

Figure 30 – Railway bridge in France (left); strengthening with special bearings (right) in Cury and Cremona [156]

Two-staged ANNs were used as classifiers by Goh et al. [157]. The authors focused on the number of measurement points, especially when the mode shape is used as an indicator in SDD applications. Increasing the number of measurement points would enhance the reliability of the ANNs while it is not feasible to install more sensors on any structure. With this motivation, the authors used ANNs for predicting the mode shape per limited number of sensors, comparing the procedure results to Cubic Spline interpolation. Within the two-staged ANNs, the mode shapes are predicted in the first stage and the SDD operation took place in the second stage. A parametric study was run to test the sensitivity of the number of measurement points and ANNs resulted in more reliable values than the CS method. FE model of a two-span reinforced concrete slab was used for simulations. As a result, the two-staged ANN was capable of predicting the mode shapes and locating the damages on the slab. This methodology needs to be verified on a real structure so that it can be generalized and used on other applications.

Yeung and Smith [158] studied SDD of a bridge using ANNs for pattern recognition of vibration signatures using a FE model of a suspension bridge. In the FE model, the traffic excitations were modeled as a series of force impulses acting along the span length representing a 40 kN vehicle travelling with various speeds. The damage based on the loss of continuity of riveted girder connections was simulated in the model. A feature vector was formed using the information from pairs of nodes and mode shape peaks in an attempt to merge sensitivity with consistency. The sensitivity of ANNs was assessed by variations on the thresholds and the level of noise added to the data. Vibration signatures obtained from the undamaged and damaged bridge structure were used to train and test the performance of two unsupervised ANNs: Probabilistic Resource Allocating Network (PRAN) by Roberts and Tarassenko [159]; and DIGNET by Thomopoulos et al. [160] and Wann and Thomopoulos [161]. As a result the SDD performance was found satisfactory with 70% identification rate when a moderate amount of noisy data; however, the methodology needs to be validated on a real structure so that it can be generalized for other applications.

For a SDD study on a cable-supported bridges using modal frequency data and Probabilistic Neural Network (PNN), Zhou et al. [162] focused on simulated noisy modal data for a suspension bridge (Tsing Ma) and a cable-stayed bridge (Ting Kau Bridge) located in Hong Kong (Figure 31). The PNN methodology uses only the modal frequency information for SDD by which a discrete number of pattern classes were used as training samples to develop a three-layer PNN for localizing damage. The training samples for all pattern class are obtained by using the computing the 3D FE model modal frequency changes. The numerical results for the TMB show that the damage at deck members can be located with high confidence when the noise level is less than 0.2. For TKB, when the noise level is less than 0.1, the SDD and localization predictions are found satisfactory. The PNN approach was found promising for SDD in noisy conditions since it describes measurement data in a Bayesian probabilistic framework. The methodology needs to be verified by real data collected from a real structure so that the procedure can be generalized for other applications.

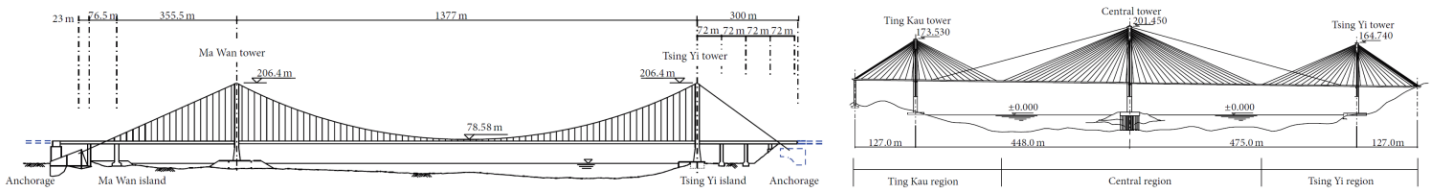

Figure 31 – Elevations of Tsing Ma Bridge (left) and Ting Kau Bridge (right) of Hong Kong used in Zhou et al. [162].

Lee et al. [163] presented a study on ANN based SDD for bridges considering errors in baseline FE models. They used an analytical beam model; an FE model of a laboratory structure and field data collected by monitoring a bridge. The ANN based SDD method utilized modal properties considering the modeling errors in the baseline FE model from which the training patterns are formed. The mode shape variations of pre-post damage conditions were used as inputs to the ANNs to minimize the modeling errors from which the training information is to be created. Laboratory test results confirm the applicability of the proposed method for various damage cases; the authors stated that the ANNs can be effectively used for SDD of the bridges subjected to traffic loadings considering the modeling errors. Moreover, for the experimental study conducted on a real bridge revealed that the SDD on a substructure was accurate for damage locations for all damage cases; however, there were minor false alarms at various locations. The procedure needs to overcome the false alarms and verified on a large-scale structure before it can be generalized for other applications.

Below, the parametric ML-based SDD approaches are summarized in Table 2 based on the extracted features, classification methods, test structures, applied excitations, and types of damage. This overall summary also specifies whether the method was used only for damage identification or for both damage identification and localization.

Table 2. Review of ML-based parametric damage detection methods. In the second column, "**A**" indicates that the study was conducted **A**nalytically, while "**E**" indicates that it was conducted **E**xperimentally. In the last column, "**I**" indicates that the method was used for damage **I**dentification and "**L**" indicates that it was used for damage **L**ocalization.

| Ref | A/E | Test structure | Type of damage | Excitation | Extracted features | Classifier | I/L |
|---|---|---|---|---|---|---|---|
| [139] | A | 5-DOF analytical model | Simulated stiffness reduction | Free vibration | Natural frequencies and mode shapes | Fuzzy neural network (FNN) | I&L |
| [140] | A | Analytical truss model | Simulated stiffness reduction | Free vibration | Natural frequencies and mode shapes | Multilayer feedforward ANN | I&L |
| [141] | A | 5-DOF analytical model | Simulated stiffness reduction | Free vibration | Natural frequencies and mode shapes | Multilayer feedforward ANN | I&L |
| [142] | A | FE model of a frame structure | Simulated stiffness reduction | Random excitation | Natural frequencies and mode shapes | Multilayer feedforward ANN | I&L |
| [143] | E | FE model of a frame structure | Simulated changes in mass and stiffness | Free vibration | Natural frequencies and mode shapes | Ensemble of two ANNs | I&L |
| [144] | A | Analytical beam and frame models | Simulated stiffness reduction | Free vibration | Natural frequencies and mode shapes | Multi-stage ANN | I&L |
| [146] | A | Analytical beam model | Simulated stiffness reduction | Free vibration | Model shapes | Multilayer feedforward ANN | I&L |
| [147] | E | Simple supported I-beam | Grinding slots in the flange | Random shaker excitation | Mode shapes | Ensemble of ANNs | I&L |
| [148] | E | 3-story frame structure | Saw cuts in the columns | Ambient vibration | Natural frequencies and mode shapes | Multilayer feedforward ANN | I&L |
| [149] | E | Beam, plate, and shell | Saw cuts | Impact hammer | Wavelet transform of the mode shapes | Multilayer feedforward ANN | I&L |
| [150] | A | FE model of a frame structure | Various | Ambient vibration | Natural frequencies, mode shapes and Ritz vectors | Multilayer feedforward ANN | I&L |
| [151] | A | 7-DOF analytical model | Simulated stiffness reduction | Free vibration | Modal parameters discretized by K-means clustering | Probabilistic neural network (ANN) | I&L |
| [152] | A | 10-DOF analytical model | Simulated stiffness reduction | Random excitation | Natural frequencies and mode shapes | Fuzzy neural network (FNN) | I&L |
| [153] | A | 7-DOF analytical model | Simulated stiffness reduction | Free vibration | Mode shapes | Fuzzy neural network (FNN) | I&L |
| [155] | E | 1-Mass-spring system. 2-Rectangular beam. | 1-Stiffness reduction 2-Saw cuts in the beam | Random shaker excitation | Natural frequencies | OS-ELM | I&L |
| [156] | A&E | 1- FE beam model 2- Field data collected by monitoring a bridge | 1- Simulated stiffness reduction 2- Data collected before strengthening was considered as damaged data | Ambient vibration | Natural frequencies and mode shapes | 1- Bayesian decision tree (BDT) 2- ANN 3- SVM | I&L |
| [157] | A | FE model of an RC slab | Simulated stiffness reduction | Free vibration | Natural frequencies and mode shapes | Two-stage ANN | I&L |
| [158] | A | FE model of a bridge | Simulated stiffness reduction | Moving vehicle load | Natural frequencies | Two Types of Unsupervised NNs (PRAN and DIGNET) | I |
| [162] | A | FE model of a cable supported bridge | Simulated stiffness reduction | Free vibration | Natural frequencies and mode shapes | Probabilistic neural network (ANN) | I&L |
| [163] | A&E | 1- Analytical beam model 2- FE model of a lab structure 3- Field data collected by monitoring a bridge | Saw cuts in the girders | Not mentioned | Natural frequencies and mode shapes | Multilayer feedforward ANN | I&L |

## 4.2 Machine-Learning methods for nonparametric vibration-based damage detection

Machine-Learning (ML) algorithms are extensively utilized to develop new nonparametric vibration-based structural damage detection methods. Traditional ML algorithms utilize the outputs of certain signal processing methods which extract damage-sensitive features without the need of modal identification. The extracted features are then fed into a ML algorithm which performs the SDD and damage localization. A wide range of feature/classifier combinations have been investigated in ML-based nonparametric SDD literature.

In a numerical study involving ML-based SDD, Chun et al. [164] presented a simplified nonparametric approach to monitor corrosion-induced thickness reduction of girders in bridges (Figure 32). The corrosion is caused by the blown anti-freezing agent from the neighboring highway bridge. The proposed SDD method is essentially an ANN based damage severity quantification method (categorized as level 3 by Rytter's classification). The feature extraction process was conducted in a nonparametric manner simply by computing the maximum and variance values of the measured acceleration signals. These values were then processed by an MLP to evaluate the integrity of the structure. FE simulations were created to produce the information required for the training and testing of the ANNs. The MLP has a single hidden layer with 20 hidden neurons. The accuracy of the proposed approach was evaluated by leave-one-out cross validation procedure (Figure 32) and found successful. By the FE model, maximum stress prediction accuracy was also verified. Since the methodology was verified only numerically, a generalization cannot be made on the use of it for other applications. This was why the authors suggested a future work on validating the model on a real bridge.

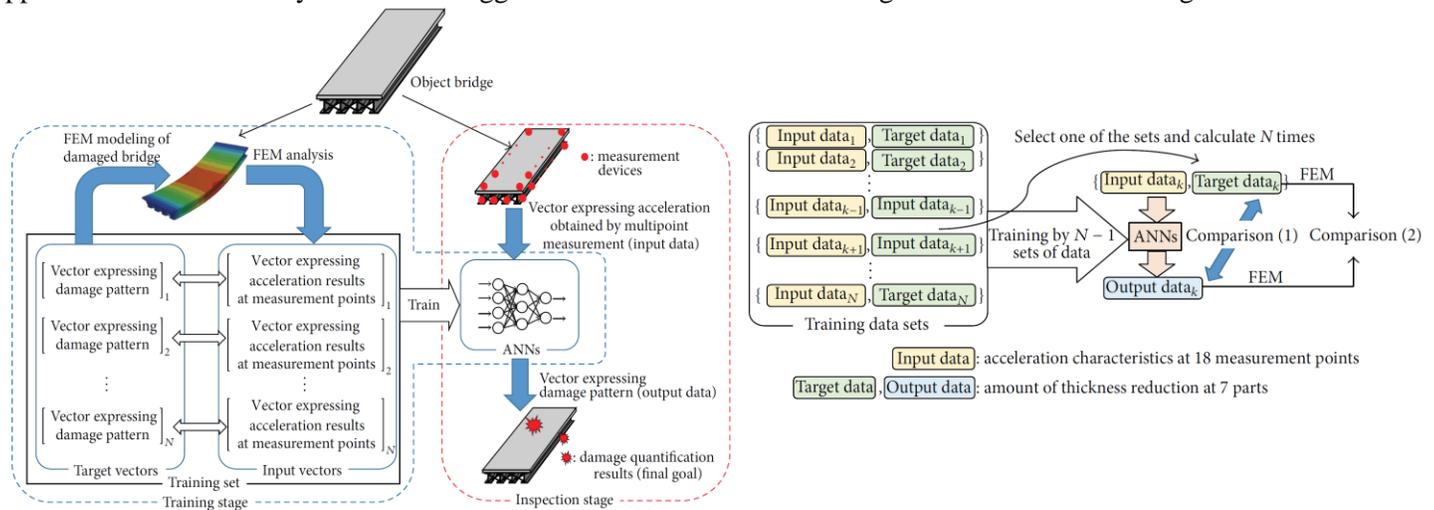

Figure 32 – Overview of the methodology (left); validation of developed method by leave-one-out cross validation (right) used in Chun et al. [164]

In time, it was observed by more researchers that simple features such as maximum and variance values of acceleration signals are not a suitable choice since they are extremely sensitive to factors other than damage (e.g. operational and environmental conditions). Based on this, researchers have started to investigate more sophisticated feature extraction techniques that are more likely to capture the characteristics of a structural damage. Most of the currently available ML-based nonparametric methods utilize AR time-series modeling for feature extraction. Figueiredo et al. [165] reported on AR modeling approach to emulate the vibration response of a small frame structure measured under random shaker excitation. A three-story laboratory specimen was excited at the base per 17 structural conditions (representative of changes in masses and stiffness). The mechanism was designed to represent the simulation in which the cracks close and open under various excitations. The identified coefficients of the AR models were utilized as damage features. The resulting data samples were used to train an Auto-Associative Neural Network (AANN) to carry out the feature classification process for environmental and operational variations. The AANN structure included three hidden layers (bottleneck, mapping and de-mapping layers). Additionally, the study investigated other techniques including Singular Value Decomposition (SVD), Mahalanobis Squared Distance (MSD) and Features Analysis (FA). All four methods were observed to be robust to distinguish damaged cases from undamaged cases; however, the comparison revealed that the AANNs resulted in better performance when compared to the other classifiers when it comes to identifying nonlinear correlation among the extracted features. The uniqueness of this work is simply the comparison of four robust algorithms to separate structural conditions with changes resulting from damage from changes caused by environmental and operational variations. The generalization and applicability of this methodology to other structures will require that new data is trained corresponding to the operational and environmental conditions specifically for the application.

A comparative study on Kernel-based approaches for SDD under varying operational and environmental conditions was conducted by Santos et al. [80] in which AR coefficients extracted from the measured acceleration signals with several classifiers including SVM, Support Vector Data Description (SVDD) Greedy Kernel Principal Component Analysis (GKPCA) and Kernel Principal Component Analysis (KPCA). The summary of the methodology is depicted in Figure 33. Damage indicators from the test data are classified based on threshold value per 95% cut-off value over the training data. Classifiers were tested using the vibration response of a three-story steel laboratory structure subjected to random shaker excitation. A column from the third floor is suspended to interact with a bumper installed on the second floor for the simulation of a nonlinear fatigue crack damage, which might open and close when subjected to various excitations. Several types of structural damage were investigated including linear damage (i.e. stiffness and mass changes) and nonlinear damage induced by boundary condition variations. Authors noted that the kernel-based algorithms revealed good performance on data groups within 2.6% to 3.4% error range. One-class SVDD and SVM are reported to be minimizing false-negative damage indications. Meanwhile, the KPCA and GKPCA algorithms are observed to be a better fit for minimizing false-positive damage indications without increasing the false-negative damage indications. Based on this, the KPCA and GKPCA were reported to display more efficient generalization performance. It is pertinent to note that this study was only concerned with SDD without localization concerns. Based on data normalization for SDD, this study compared ML algorithms to set the normal condition as a function of operational and environmental changes. An important observation is that these ML methods do not require a direct measure of the sources of variability, yet they rely only on measured response time-series data collected under varying operational and environmental effects.

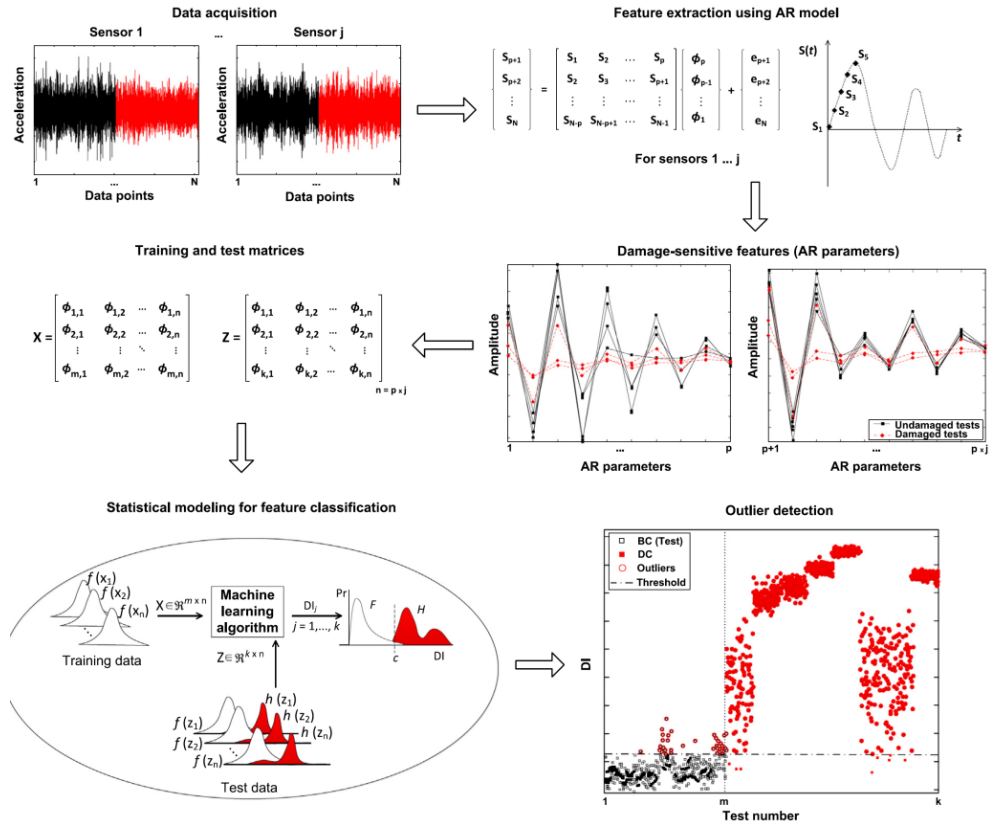

Figure 33 – Overview of the SDD architecture in Santos et al. [80]

In a recent study on data-driven SVM with optimization techniques, Gui el al. [166] used AR modeling to extract sixteen damage sensitive features from a set of experimental vibration data for SDD on a three story laboratory frame. Three optimization techniques, namely Particle Swarm Optimization (PSO), Grid Search (GS), and Genetic Algorithms (GAs) were utilized to tune the hyperparameters of the SVM. The cross-validation approach is utilized to reach the most reasonable parameters (Figure 34) during data prediction in an attempt to avoid the over-fitting for the SVM model. The study focuses on performance comparisons of the optimization-based ML methods (GA+SVM, PSO+SVM, GS+SVM) with two types of damage sensitive features: Residual Error (RE) and Auto-Regressive (AR) for SDD. An SVM was trained and used to carry out the SDD and localization process. The results showed that combinations of GS+RE, GA+RE, PSO+AR, PSO+RE all resulted in excellent classification performances. RE based SVM revealed 100% accuracy in all cases which means RE is very sensitive to nonlinear behavior caused by damages. On another note, among the three methods, GS method was reported to have the best accuracy on data classification thanks to its independent parameters option. The generalization of this SDD procedure for other structures will require verification on a large-scale structure.

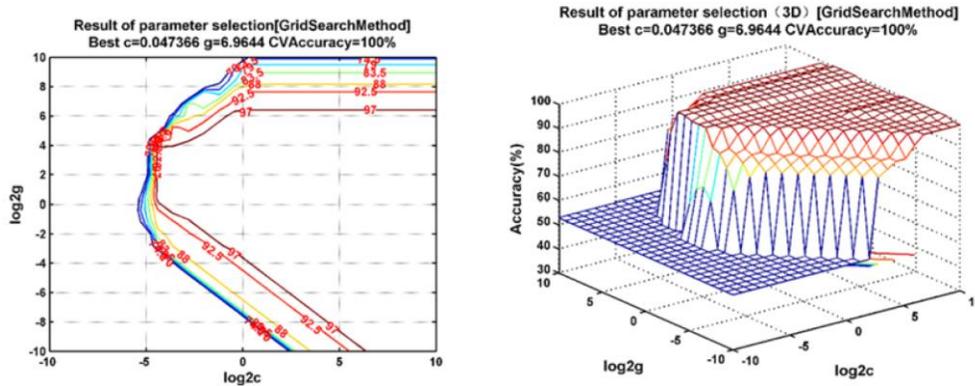

Figure 34 - Parameter Selection for Residual Error using the Grid Search Technique in Gui el al. [166]. 2D (left) and 3D (right) contours.

In addition to the aforementioned methods which rely exclusively on AR coefficients, there are studies which used AR modeling in conjunction with other techniques in an attempt to refine the extracted features. For instance, Lautour and Omenzetter [167] utilized AR modeling to extract a large number of features through the vibration response of a three-story laboratory frame and a benchmark structure. AR models were used simply to fit the acceleration response of the two experimental structures (Figure 35). The AR model coefficients were considered to be damage-sensitive features and used as input into an ANN. Principle Component Analysis (PCA) was then utilized for dimensionality reduction of the extracted feature vectors. Using the reduced dataset, a MLP was trained to carry out the feature extraction process. For classifications, it was reported that single hidden layer MLPs with varying numbers of hidden neurons were attempted; and as a result, a single hidden layer ANN with four hidden neurons was found to be the optimum. This architecture was observed to produce the best outcome with 3.4% of classification errors among a sample of 88 data points. On the other hand, for the estimation of remaining stiffness, another MLP with single hidden layer and five hidden layer neurons was found to be the best. As a conclusion, the tests on the laboratory structure revealed that for stiffness reductions of 7%-10%, the MLP was successfully classifying the damage with 97% accuracy. As a result of this study it can be stated that the combination of AR models and ANNs are accurate tools for SDD even though small number of damage-sensitive features and limited sensors are utilized. Considering the success of the methodology on both laboratory and benchmark structures, the approach can be used on other applications.

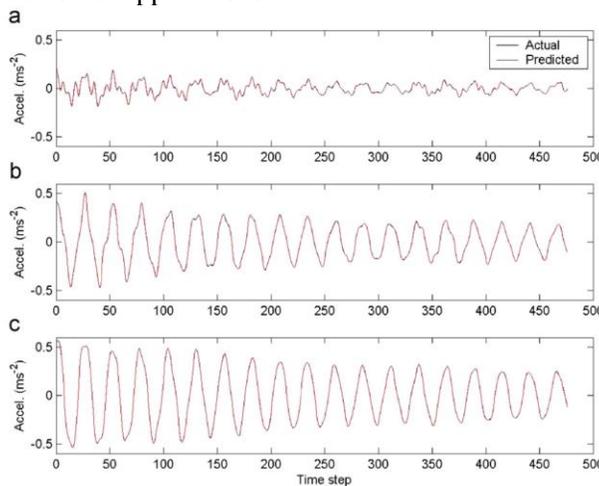

Figure 35 – Perfect match of actual and predicted acceleration responses for the laboratory frame via AR model (a) story 1, (b) story 2 (c) story 3 in Lautour and Omenzetter [167].

On another ML-based nonparametric SDD study, Dackermann et al. [168] applied PCA directly over the raw vibration signals in time-domain for feature extraction without utilizing AR modeling. The methodology takes advantage of a system based on a combination of PCA and ANNs in an attempt to resolve primary issues regarding incomplete data sets, measurement noise and limited number of sensor arrays. The output of PCA, referred to as PCA-compressed damage indices (DIs), were fed into an ensemble of ANNs trained for SDD and localization. Numerical and experimental investigations conducted on a simply supported experimental beam demonstrated the noise-filtering ability of the PCA. In the FE model, various levels of Gaussian noise are introduced to the numerical data. Based on the findings of this study, it was noted that for noise levels of 1% and 2%, only one or two very light damage states were incorrectly predicted; meanwhile for noise levels of 5% and 10%, all medium and severe damage states were correctly predicted. On another note, for quantifying numerically simulated damage states, the samples trained with 1%, 2% and 5% noisy data resulted in accurate SDD for all damage states; whereas the samples trained with 10% noisy data inaccurately quantified one very light and two light damage state scenarios. While the methodology was tested on a determinate small-scale laboratory beam, it would have made more sense to generalize this approach after testing it on a large-scale structure.

Bandara et al. [169,170] proposed a slightly different damage detection strategy in which PCA was applied on the frequency response functions (FRFs). A 10-story FE model was studied in [169] and stiffness reduction was simulated. For [170], a three story laboratory frame structure with adjustable bumper and the suspended column was studied (Figure 36). In [169], the FRFs were estimated from the vibration response in the time-domain recorded by an array of accelerometers placed on the 10-story building. The damage indices extracted by PCA were used to train a MLP (multilayer feedforward ANN), which was responsible for localizing and quantifying structural damage on the building. The initial ANN architecture used in the study includes four layers: one input layer, two hidden layers and one output layer. To avoid over-fitting and to maximize the generalization capability, a validation data set is utilized by dividing the available batch into 50%, 25% and 25% portions for training, validation and testing purposes, respectively. Meanwhile, for 60 damage cases (3 damaged floors × 5 columns at each floor × 4 severity conditions) and 4 noise levels, there are 240 data sets for ANN training. Furthermore, for the identification of the damaged floor, there are 13 ANNs representing 12 ANNs with one additional ANN from addition of FRFs. Also, one input layer of 11 nodes, two hidden layers with 7 and 4 nodes; and an output layer with three nodes are utilized for the ANN for the SDD of the floor. As a result, the damage indices indicate that the multi-stage ANNs described in the study can precisely identify damages with small severity, confirming the accuracy of proposed method on transforming FRFs to damage indices. The methodology is almost the same in [170] for the laboratory structure. An example for comparison of FRFs for SDD is shown in Figure 36. The results revealed that the different levels of nonlinear damage can be detected successfully by the trained ANNs. Also, ANNs trained with summation of FRFs resulted in higher accuracy for SDD when compared to ANNs trained with individual FRFs only. Even though it was reported that the methodology used in [169] and [170] was suggested for a real structure; a verification on a large-scale structure would still be needed for the generalization of this method.

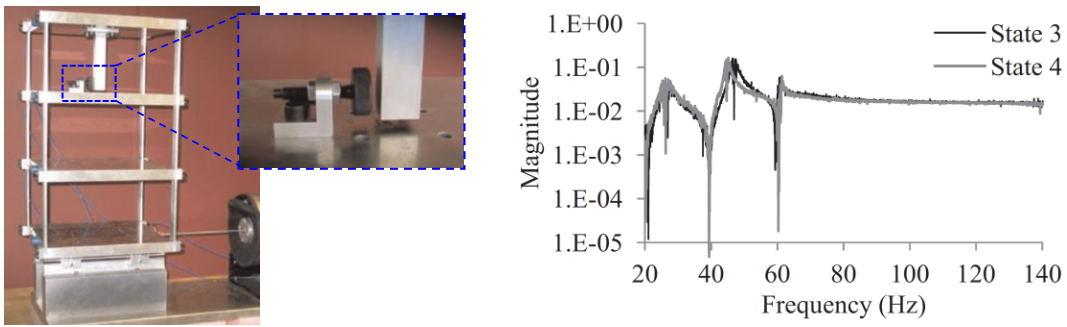

Figure 36 – Laboratory structure (left) and example FRF comparison for SDD (right) from Bandara et al. [170]

For ML-based nonparametric SDD, very few studies have utilized feature extraction techniques other than AR modeling and PCA. For instance, Liu et al. [171] used wavelet decomposition on the acceleration signals for feature extraction, and used a typical MLP for classification. Utilizing the existing data of a benchmark [129] the SDD methodology is a combination of Wavelet Packet Transform (WPT), multi-sensor feature fusion and ANN pattern classification. The accelerations were decomposed using orthogonal wavelet and the relative energy of decomposed frequency band was calculated. Then, the input feature vectors of ANN classifier were built by fusing wavelet packet relative energy distribution of sensors. As a conclusion, the combination of the WPT, multi-sensor feature fusion and the ANN model was found to be promising for SDD and localization; however, the approach needs to be verified on an existing structure before it can be generalized for various other applications.

Similarly, Ghiasi et al. [172] introduced a nonparametric ML-based method integrating Wavelet Packet Decomposition with an SVM classifier using the benchmark data of [129] and a dome truss, as numerical examples. Thin plate spline wavelet kernel function of this approach is integrating Thin Plate Spline Radial Basis Function kernel with local characteristics and a modified Littlewood–Paley function. Parameters were optimized by a social harmony search algorithm for SDD and the results were compared to the least square SVM based model on other kernels. It is found that Thin Plate Spline Littlewood–Paley wavelet kernel has better accuracy and learning ability than approaches utilizing conventional kernels. Being tested on a benchmark data of a laboratory structure and an FE model of a second structure, the method is promising for other SDD applications. A real-time SDD application on in-service structures (which is suggested as a "future work" by the authors) would validate the generalization of this method for other applications.

In another nonparametric ML-based system identification and response prediction study, Zhu et al. [173] used interval modeling technique for feature extraction along with an Adaptive Neuro-Fuzzy Inference System (ANFIS) for SDD and damage classification, again, utilizing the benchmark data of [129]. The FE model of the benchmark structure was created in Abaqus software in which the damage is introduced as element removals. Per the noise-polluting displacement response data based on the hammer excitation, the method was able to detect damage within 0.02 to 0.03 seconds. After a proper training, ANFIS outputs a displacement vector. With 10% noise used in training the ANFIS model produced successful SDD results. Despite the success of this approach, a validation on a real structure is needed so that the methodology can be generalized for other applications.

Another example is the nonparametric ML-based method proposed by Abdeljaber and Avci [174]. The method involves training several Self-organizing Maps (SOMs), which are known to be a member of unsupervised classifiers, to extract a number of meaningful damage indices (Figure 37) from vibrations response of a structure measured under random excitations [174,175]. A MLP was then trained to process the extracted damage indices in order to conduct SDD and determine the severity of structural damage. This method was verified analytically using a FE model of a steel grid structure over damage cases involving simulated stiffness loss (Figure 37) as well as boundary condition changes. The authors used similar versions of the same approach in other studies ([45] and [108]) and achieved successful SDD and localization. Even though the algorithm was shown to be successful with noisy signals in the analytical models, an experimental validation on a real structure is needed before the methodology is generalized for other applications.

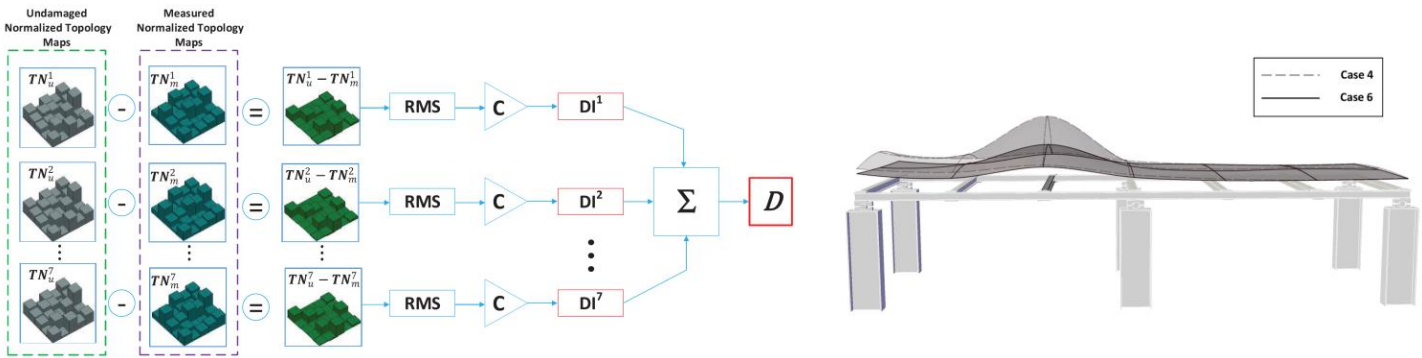

Figure 37 – Damage assessment methodology (left) and an example comparison of damage index distribution (right) in Abdeljaber and Avci [174]

At the end of this section, the nonparametric ML-based SDD approaches are summarized in Table 3 based on the extracted features, classification methods, test structures, applied excitations, and various types of damage. This overall summary also indicates whether the method was used only for damage identification or for both damage identification and localization.

Table 3. Review of ML-based nonparametric damage detection methods. In the second column, "**A**" indicates that the study was conducted **A**nalytically, while "**E**" indicates that it was conducted **E**xperimentally. In the last column, "**I**" indicates that the method was used for damage **I**dentification and "**L**" indicates that it was used for damage **L**ocalization.

| Ref | A/E | Test structure | Type of damage | Excitation | Feature extraction method | Classifier | I/L |
|---|---|---|---|---|---|---|---|
| [164] | A | FE model of a bridge | Reduced thickness of structural members | Random excitation | Maximum and variance of acceleration signals | Multilayer feedforward ANN | I&L |
| [165] | E | Steel frame structure | Changes in mass and stiffness | Random shaker excitation | Autoregressive (AR) modeling | 1- ANNs<br>2- Factor analysis<br>3- Mahalanobis distance<br>4- Singular value decomposition | I |
| [80] | E | Steel frame structure | Loosening of base-to-column connections | Random shaker excitation | Autoregressive (AR) modeling | 1- one-class support vector machine<br>2- support vector data description<br>3- kernel principal component analysis<br>4- greedy kernel principal component analysis | I |
| [166] | E | Steel frame structure | Loosening of base-to-column connections | Random excitation | Autoregressive (AR) modeling | Support vector machine (SVM) | I&L |
| [167] | E | Steel frame structure | Various | Random shaker excitation | Autoregressive (AR) modeling followed by principle component analysis (PCA) | Multilayer feedforward ANN | I&L |
| [168] | E | Simple supported beam | Saw cuts | Impact excitation | Principle Component Analysis (PCA) | Ensemble of ANNs | I&L |
| [168] | A&E | Steel frame structure | Joint damage | Random shaker excitation | Principle Component Analysis (PCA) applied on the FRFs | Ensemble of ANNs | I&L |
| [169] | A | FE model of a high-rise building | Simulated stiffness reduction | Random excitation | Principle Component Analysis (PCA) applied on the FRFs | Multilayer feedforward ANN | I&L |
| [170] | E | Steel frame structure | Loosening of base-to-column connections | Random excitation | Principle Component Analysis (PCA) applied on the FRFs | Multilayer feedforward ANN | I&L |
| [171] | A | 12-DOF analytical model | Simulated stiffness reduction | Random excitation | Wavelet decomposition | Multilayer feedforward ANN | I&L |
| [172] | E | Steel frame structure | Various | Random shaker excitation | Wavelet decomposition | Support vector machine (SVM) | I&L |
| [173] | A | FE model of a steel frame | Elements removal from the FE model | Free vibration | Interval modeling | Adaptive neuro-fuzzy inference system (ANFIS) | I&L |
| [174] | A | FE model of a grid structure | 1- Simulated stiffness loss.<br>2- Simulated changes in boundary conditions | Random excitation | Self-organizing maps (SOMs) | Multilayer feedforward ANN | I&L |

## 4.3 Drawbacks and Limitations

In the earlier subsections, various approaches of ML algorithms in both parametric and nonparametric vibration-based SDD applications are described and discussed. A particular focus is drawn onto feature extraction and feature classification techniques which are utilized to perform the SDD operation. From the comprehensive literature review presented so far, it can be concluded that in the SDD context both approaches require extracting a fixed set of hand-crafted (i.e. user-defined) features from the vibration signals. Accordingly, it is reasonably expected that the successful performance of ML based SDD methods significantly relies on the extracted features and the classifier choices. From this perspective, selection of extracted features is so crucial that the characteristic information of the analyzed signals can be captured effectively. In addition to the extracted feature type, a convenient classifier is required for an accurate and robust damage detection and localization.

As observed in both parametric and nonparametric vibration-based SDD applications of ML algorithms reviewed so far, researchers in this field have examined many feature/classifier combinations by trial-and-error in search of the best combination capable of optimally characterizing the structural damage. However, there are several issues associated with the currently available ML-based parametric and nonparametric SDD methods that can be summarized as the following:

1) There is no guarantee that a specific feature/classifier set will be the best choice for all types of SDD exercises. Put another way, a particular feature/classifier set observed to be optimum for a certain type of structure might not be a reasonable choice for any other type of structure.
2) Similarly, there is no guarantee that a certain feature/classifier set will be the most optimum choice for all types of structural damage. For example, a specific feature/classifier set which is observed to be a good fit for identifying stiffness loss might not able to detect changes in boundary conditions.
3) Using a fixed set of hand-crafted features or unreasonable classifiers would probably bring forth unsuccessful SDD performance.
4) Feature extraction techniques such as modal estimation, AR modeling, and PCA usually cause considerable computational complexity and time, which hinder the use of ML-based methods in real-time SHM operations.
5) Most parametric and nonparametric ML-based methods, with very few exceptions, are centralized. This means that all measured signals should be synchronized and transferred to a single processing unit before carrying out the damage detection process. Yet, the centralized methods have requirements which are sometimes difficult to satisfy. For example, all sensors in the network need to remain fully functional for the operation to be successful. Also, the centralized approaches are difficult to implement in wireless sensor network (WSN) applications.

## 5. Vibration-based structural damage detection by Deep-Learning

As discussed earlier, conventional ANNs rely on shallow nets consisting of an input, one or two hidden layer(s), and an output layer. A neural network including more than three layers can be counted as "deep" learning network. In other words, a Deep Neural Network is basically an ANN housing multiple layers between the input and output layers. Deep Learning (DL) is the most recent accomplishment of the Machine Learning age. Having become a research hotspot, DL is actively being utilized in our daily lives, providing solutions to so-called "very difficult" problems of the previous decade. Even prior to the launch of the AlexNet [176], it can be argued that the DL era has begun in 2006 with the article published by Hinton and Salakhutdinov [177]. In this article the authors described the effect of "the depth" of an ANN in ML, indicating that ANNs with several hidden layers can have a powerful learning ability. This ability can even be enhanced with the increasing depth (and the number) of hidden layers, resulting in with the term, "Deep" Learning. In other descriptions, DL is simply described as a ML branch which has the capability to manage complex patterns and objects in very large size datasets. In a broader sense, the main drawback of the conventional ML based methods is the fact that they are highly dependent on hand-crafted features. Such features might be sub-optimal, which means that they cannot accurately characterize the measured signal. As such, when a classifier is trained based on sub-optimal features, it tends to result in unsatisfactory classification performance, and therefore unreliable SDD results, e.g.,[178]. Most DL methods and in particular deep CNNs can learn to extract "optimized" features directly from the raw data for the problem at hand to maximize the classification accuracy. This is indeed the key characteristic for improving the classification performance significantly which made them attractive to complicated engineering applications.

In DL, multiple abstract layers communicate with each other. In other words, DL allows computational models that are composed of multiple processing layers to learn representations of data with several levels of abstraction. DL algorithms can learn not only to correlate the features to the desired output, but also to run the entire feature extraction process. Therefore, without the need to extract features in advance, a DL system with proper training can find the direct mapping from the raw inputs to the final outputs. This enables DL algorithms excel in complex tasks by breaking them down into smaller and simpler parts.

DL architectures can be categorized into four major groups as per Patterson and Gibson [179]: Unsupervised Pretrained Networks (UPNs); Convolutional Neural Networks (CNNs); Recurrent Neural Networks; and Recursive Neural Networks. There are three further architecture types under UPNs: Autoencoders (also known as Deep Autoencoders), Deep Belief Networks (DBNs), and Generative Adversarial Networks (GANs) as shown in Figure 38. Among the DL architectures, the optimum architecture typically depends on the type of the application. Since the UPNs (Autoencoders) and CNNs are the only DL tools that have so far been utilized in vibration-based SDD applications; publications regarding these two are included in this review.

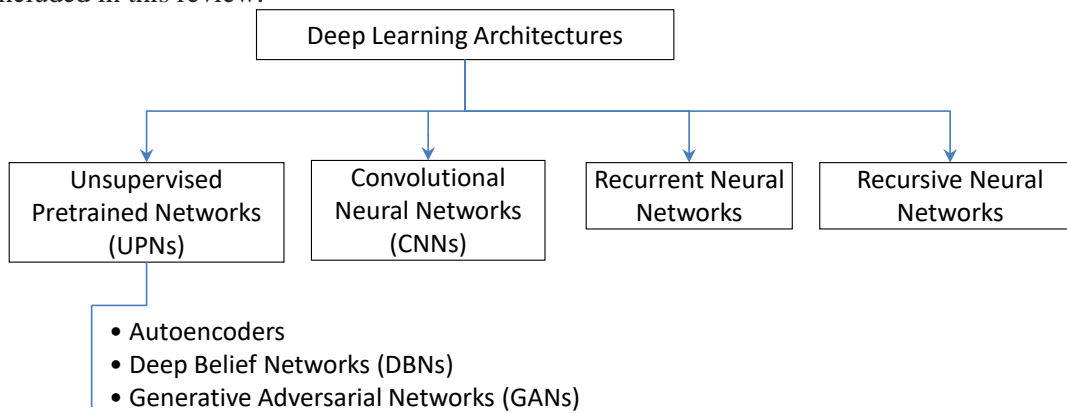

Figure 38 – Deep-Learning architectures

### 5.1 Unsupervised Pretrained Networks (UPNs) and Autoencoders

The architectures under UPNs are Autoencoders, Deep Belief Networks (DBNs), and Generative Adversarial Networks (GANs). Among the UPN architectures, Autoencoders is the only architecture used in vibration-based SDD applications; therefore they are covered and discussed in this paper. The efficiency of the Autoencoders against conventional ANNs is presented by Hinton and Salakhutdinov [177]. Autoencoder is basically a technique to find fundamental features representing the input images. A simple autoencoder will have one hidden layer between the input and output, while a "Deep Autoencoder" will have multiple hidden layers (the number of hidden layer varies). Yet, often in DL terminology, the term Autoencoder is used for "Deep Autoencoder". In Autoencoders, the learned features describe the original data in a much better way, which makes them an excellent choice for classification. Autoencoders have not been used in SDD until Fallahian et al. introduced the ensemble classification method based on weight majority voting [180] and the accompanied work [181]. Fallahian et al. referred to the Autoencoder method described in Hinton and Salakhutdinov [177] and they called it Deep Neural Networks (DNNs). While this terminology is not wrong, based on the categorization of Patterson and Gibson [179], in this paper Fallahian et al. work is discussed under Autoencoders.

In Fallahian et al. [180], the motivation of the authors is the fact that in vibration-based SDD methodologies where modal properties are used as damage indicators, it was reported multiple times that these properties are sensitive to damage as well as they are for temperature variations. Therefore, they proposed a new SDD algorithm to examine the state of a structure in the presence of uncertainties such as temperature and noise. This new SDD methodology is based on an integrated system of two strong pattern recognition methods employed for SDD: Autoencoder and Couple Sparse Coding (CSC). PCA is also employed for decreasing the dimension of the measured FRF data and compress the dataset to build distinguishable patterns. For the SDD task, it is assumed that the elastic moduli of the materials are temperature-dependent while the temperature is treated as an input parameter for the system. Using the data of a truss bridge and experimental data of the I-40 bridge, the success of the proposed SDD technique is verified in the presence of temperature changes, noise, measuring and modeling errors. The authors highlighted that the temperature needs to be recorded at as many locations as possible because the temperature gradients may potentially affect the precision of the results.

In a follow-up study [181], Fallahian et al. presented the Autoencoder–CSC ensemble this time with experimental validation on data from an aluminum beam in [51]. Two damage scenarios were introduced on the beam; the first one was simply a single cut and the second one was a double cut resulting in various stiffness reductions at the cross section. Based on the impact hammer excitations for damaged and undamaged conditions, the pattern recognition-based SDD technique is run via the FRF of the structure per the Autoencoder and CSC classification. The process was also repeated and verified with a 3D FE model of a truss bridge. The proposed ensemble method is again found effective in the presence of environmental variations and various other uncertainties compared to the performances of each of the individual methods when used alone.

On another study by Pathirage et el. [182], the authors introduced another DL method for SDD based on autoencoders highlighting the dimensionality reduction and relationship learning (Figure 39). The system performs pattern recognition between the modal and structural stiffness parameters. The first task in the paper was reducing the dimensionality of the original input vector and preserving the necessary information through multiple hidden layers. The following task was running the relationship learning between the reduced features and stiffness reduction parameters. Modal characteristics, such as natural frequencies and mode shapes, are used as the input and the structural damage are considered as the output vector. A pre-training is used to train the hidden layers in the autoencoders layer by layer, and fine tuning is done to optimize the whole network. The methodology was verified on numerical and experimental models based on steel frame structures; and more efficient results are obtained when compared to the traditional ANN methods (Figure 39). As for future work, the authors stated that the methodology will be expanded to include other modal characteristics such as FRFs in an attempt to detect smaller damages per noisy data and other uncertainties; therefore the methodology cannot be generalized for other applications yet.

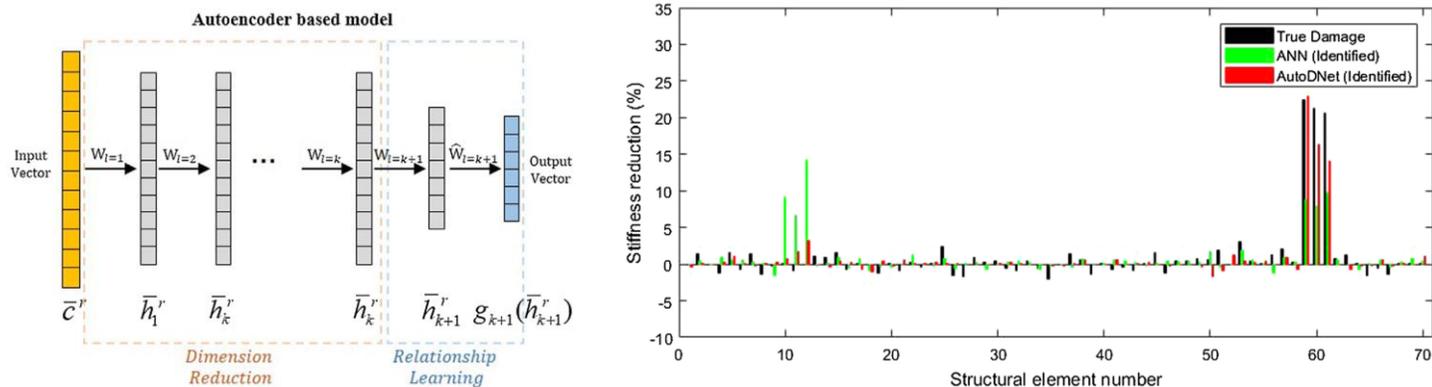

Figure 39 - Autoencoder Model (left); and SDD results of a multiple damage case from ANN and the Autoencoder Model for Case 4 (right) in Pathirage et el. [182].

## 5.2 Convolutional Neural Networks

Convolutional Neural Networks (CNNs) have recently become the most popular type of DL algorithms thanks to their capability of learning directly from the raw signals in a large-scale dataset. This is why nowadays numerous researchers have started to apply CNNs in order to address the aforementioned limitations of conventional ML-based SDD methods. The recent applications of CNNs in vibration-based SDD of civil structures are reviewed in this section.

CNNs are a type of supervised multi-layer feed-forward artificial neural networks (ANNs). Relatively recently, CNNs are reported to become an evident standard for DL tasks since they achieved the state-of-the-art efficiency in almost all engineering applications they have been utilized [88,89,183], outperforming alternative methods with distinctive performance difference. The structure and components of CNNs are mainly inspired by cells in the primary vision cortex of mammalian brain [184]. Initially, the CNNs have shown superior performance in Computer Vision applications such as object recognition in large image databases and face recognition [89,176]; and CNNs have later been adopted by in various other fields. The success of CNNs can mainly be attributed to the following advantages:

1. CNNs fuse the feature extraction classification and feature extraction stages into a single learning body. CNNs can learn to optimize the features during the training phase directly from the raw input.
2. Since CNN neurons are sparsely-connected with tied weights, CNNs can process large inputs with a great computational efficiency compared to the conventional fully-connected Multi-Layer Perceptrons (MLP) networks.
3. CNNs are immune to small transformations in the input data including translation, scaling, skewing, and distortion.
4. CNNs can adapt to different input sizes.

A sample CNN architecture is presented in Figure 40 where it is used to classify a 24×24-pixel grayscale image into two categories. The network consists of two convolution and two pooling layers. The production of the last pooling layer is processed through a single fully-connected layer and followed by the output layer that produces the classification output.

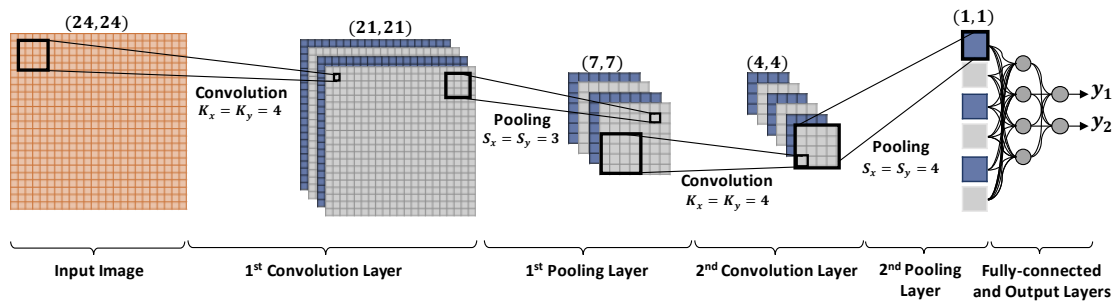

Figure 40 - The illustration of a sample CNN with 2 convolution and one fully-connected layers.

The interconnections feeding the convolutional layers are assigned by weighting filters ($w$) having a kernel size of $(K_x, K_y)$. The pooling layers are assigned by an appropriate subsampling factor $(S_x, S_y)$. In this example, the kernel sizes corresponding to the two convolution layers were set to $K_x = K_y = 4$, while the subsampling factors were chosen as $S_x = S_y = 3$ for the first pooling layer and $S_x = S_y = 4$ for the second one. Note that these values were deliberately selected so that the last pooling layer output (i.e. fully-connected layer's input) are scalars. However, the number of fully-connected layers along with the number of neurons in the convolution process, and fully-connected layers were arbitrarily chosen for illustration purposes. The output layer is composed of two fully-connected neurons in parallel to the number of classes to which the image is categorized. From this example, the following items can be listed regarding the structure and components of CNNs:

- CNNs are composed of a number of convolution and pooling layers usually, but not necessarily, superseded by fully-connected layers.
- The structural organization of the CNNs is determined by the following hyperparameters:
  o The number of convolution, pooling and fully-connected layers.
  o The number of neurons in each convolution and fully-connected layer.
  o The kernel sizes $(K_x, K_y)$ and subsampling factors $(S_x, S_y)$ of the convolution and pooling layers.

- In forward-propagation (FP), the size of the input image gets gradually reduced by successive convolution and subsampling operations. The number of convolution and pooling layers along with the selected kernel sizes and subsampling factors are usually selected in a way that the inputs to the 1st fully-connected layer are scalars.

CNNs are trained principally in a supervised manner by a stochastic gradient descent method, or by the so-called back-propagation (BP) algorithm. During each iteration of the BP, the gradient magnitude (or sensitivity) of each network parameter such the weights of the convolution and fully-connected layers is computed. The parameter sensitivities are then used to iteratively update the CNN parameters until a certain stopping criterion is achieved. A detailed description of BP in CNNs can be found in [90], [185] and [186].

### 5.2.1 Applications of CNNs for vibration-based SDD

Owing to their superior learning ability, CNNs have recently been used to develop state-of-the-art techniques for vibration-based SDD in civil structures. For instance, in a numerical study, Yu et al. [187] designed and trained a CNN for localizing and quantifying structural damage in a five-story laboratory benchmark structure. The structure by Wu and Samali [188] is employed for numerical investigation. Since conventional (deep) 2D CNNs are only able to deal with 2D data, the 1D vibration signals acquired by 14 accelerometers were converted to a 2D representation simply by concatenating the 14 measured signals into a matrix. The data required for training the CNN was gathered from a numerical model of the monitored structure under various damage scenarios. The proposed model in [187] was based on a 10-layer setup, which is composed of 1 input layer; 3 convolutional layers and 3 sub-sampling layers as shown in Figure 41. The system configuration also includes two fully connected layers and an output layer. A big kernel (1000x1 in size) is utilized in the first convolutional layer to overcome high-frequency noise, superseded by a sub-sampling layer (3x1 in size). This would form a broader feature map in the first convolutional layer via the reduced spatial resolution. As a following step, a convolutional layer with kernel (30x3 in size) and a sub-sampling layer (3x1 in size) are added in the second convolutional layer and the second sub-sampling layer, respectively. Meanwhile, a convolutional layer (10x3 in size) and a subsampling layer (3x1 in size) are used in the third convolutional layer and the third sub-sampling layer, respectively. From this point on, two fully connected layers are added. The first fully connected layer essentially transforms the output of the third sub-sampling layer into a one-dimensional feature map. The technique proposed by the authors is tested on a laboratory frame subjected to excitations of the El Centro earthquake. The results reveal that the proposed CNN model successfully detects damage on raw noisy signals and outperforms other conventional ML-based techniques. Yu et al. indicated that experimental and field studies still need to be conducted for verification purposes since only numerical investigation is performed in [187].

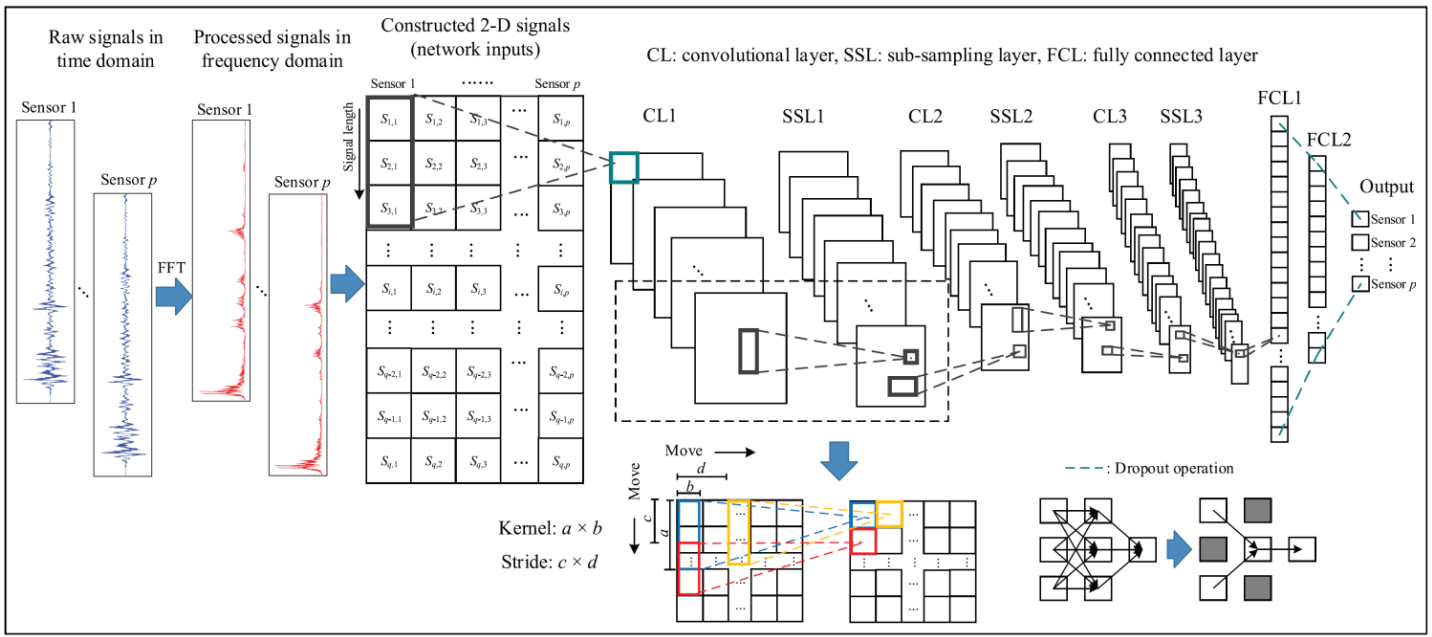

Figure 41 – The CNN based model by Yu et al. [187]

Khodabandehlou et al. [189] proposed a similar CNN-based SDD technique. A one-fourth scale laboratory structure which is an idealized two-span frame of a cast-in-place, post-tensioned, continuous reinforced concrete box-girder bridge was used to experimentally demonstrate the proposed method. The structure was used to generate vibration data corresponding to four damage levels ranging from "no damage" to "extreme damage". For each damage level, the data measured by a number of accelerometers was concatenated into a single 2D matrix. The resulting dataset was used to train a deep CNN having 5 convolutional and 4 FC layers. With 14 accelerometers placed on the test frame and with 48 measurement components, the feature matrix size becomes (610x14). As such, the feature vector is a (8540x1) in size. 2D CNNs were employed for dimension reduction. Each acceleration history was formed into a (122x70) matrix, and then fed as CNN's input image. The pooling layer computes the maximum value which is repeated five times. At this stage, the last pooling layer's output is basically reduced into a column vector and it becomes an input for the fully connected layer. When the convolution operation is repeated five times, it results in reduction on the input vector dimension from (122x70) to (4x32). Consequently, the dimension of fully connected layer input is reduced from (128x1) to (60x1). In turn, among a total of 48 shake table measurement sets, 8 were selected for the tests and 40 were utilized for CNN training. As a result, it was demonstrated that the CNN was successful at quantifying the overall condition of the bridge directly from the measured vibration response, for all damaged conditions. The method was able to capture very small changes in structural condition.

In a transmissibility based SDD study by Cofre-Martel et al. [190,191], the authors introduced a CNN-based approach for SDD and damage quantification running on images from transmissibility functions using image processing capabilities of CNNs. Structural damage was introduced as stiffness reductions which was proved to be relevant to transmissibility functions. The feature maps are fed into a MLP to perform SDD. The methodology was verified on two case studies (a mass-spring system and a structural beam) where training data are generated by calibrated FE model data contaminated by noise. The methodology resulted in satisfactory SDD performance utilizing raw vibrational data only.

The covered CNN applications for SDD so far in this section utilized conventional 2D CNNs which are naturally suitable for 2D data such as images or video frames. As an alternative, one-dimensional convolutional neural networks (1D CNNs) have recently been proposed [90] for arrhythmia detection in electrocardiogram (ECG) records and later used in many 1D signal repositories [16,90,185,186,192–197]. 1D CNNs are similar to 2D CNNs except for some architectural variations. For example, the kernels of 1D CNNs are obviously 1D arrays and 1D convolutions are performed. Recent studies [16,35,90,185,186,192–200] have shown that compact 1D CNNs can yield certain advantages over their conventional (deep) 2D counterparts when dealing with 1D signal repositories due to the following reasons:

- Rather than 2D matrix convolutions, FP and BP in 1D CNNs require simple array operations. This means that the computational complexity of 1D CNNs is significantly lower than 2D CNNs.
- Recent studies have shown that 1D CNNs with compact architectures (small number of hidden layers and small number of neurons) were able to learn challenging tasks when the training data is scarce. This is usually the case over certain problems with isolated data such as personalized ECG data repositories [90], engine vibration signals [35] or voltage/current data of a special type of power electronics [196]. This is usually not a viable option for deep CNNs since such deep networks will cause "overfitting" problem on such limited datasets.
- Usually, any operation (BP or FP) of deep 2D CNNs requires special parallelized hardware setup (e.g. GPU farms). On the other hand, typical single-CPU implementations are quite feasible for 1D CNNs.
- Due to their low computational requirements, 1D CNNs are well-suited for real-time applications especially on low-cost mobile devices.
- In [35,198], 1D CNNs have shown a superior performance in vibration-based fault detection ball bearings in rotating machinery. In particular, Abdeljaber et al. in [35] proposed a novel technique that satisfactorily detected and identified the severity of structural damages in both inner and outer ring of the bearing. Figure 42 shows an illustration for the training of the two 1D CNNs and the accompanied SDD task in order to accomplish this objective. The minimally-trained classifiers, $CNN_i$ and $CNN_o$ are reported to detect the new damage states with a precision close to 100%.

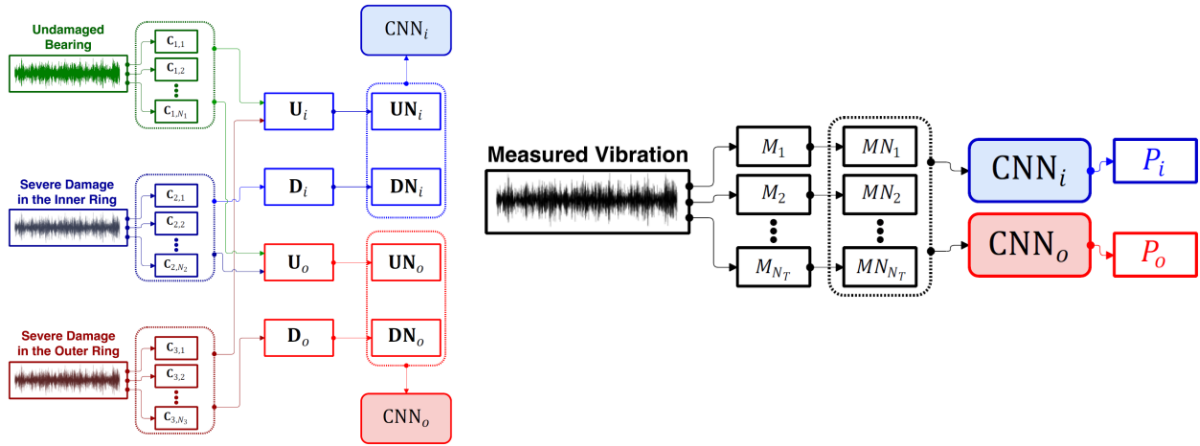

Figure 42 - Training of the two 1D CNNs, $CNN_i$ and $CNN_o$ (left); quantification and localization of bearing damage using $CNN_i$ and $CNN_o$ (right), in Abdeljaber et al. [35].

Regarding the use of 1D CNNs for vibration-based SDD, Abdeljaber et al. [185] conducted an experimental study on a relatively large laboratory frame instrumented with wired uniaxial accelerometers. The vibration response of the frame under 31 damage scenarios was measured and used to train a 1D CNN for each sensor placed at a node. Each 1D CNN was only responsible for processing the local data measured at the corresponding location. The performance of this 1D CNN-based damage detection method was tested under a large number of single and double damage cases. A complexity analysis was also conducted to estimate the computational time required for the 1D CNNs to process the measured signals. It was revealed that the method was very successful in detecting and locating structural damage in real-time. The data for study [185] is assembled and published on a public website by the authors [201]. The source code of 1D CNNs is also publically shared online [201].

In a follow up experimental study by Avci et al. [16] and [202], the 1D CNN-based method developed in [185] was integrated with a Wireless Sensor Network (WSN). The method was modified to allow it to analyze the signals measured by the triaxial wireless sensors in an attempt to determine the direction in which the damage-sensitive features are more pronounced. The modified damage detection method was tested against a number of damage states introduced to the structure shown in Figure 43. The results revealed the superior capability of the proposed technique to identify and localize damage directly from the ambient vibration response of the structure. All 1D CNNs trained in this study had a compact structure (only two CNN layers each with four neurons superseded by two MLP layers with five neurons).

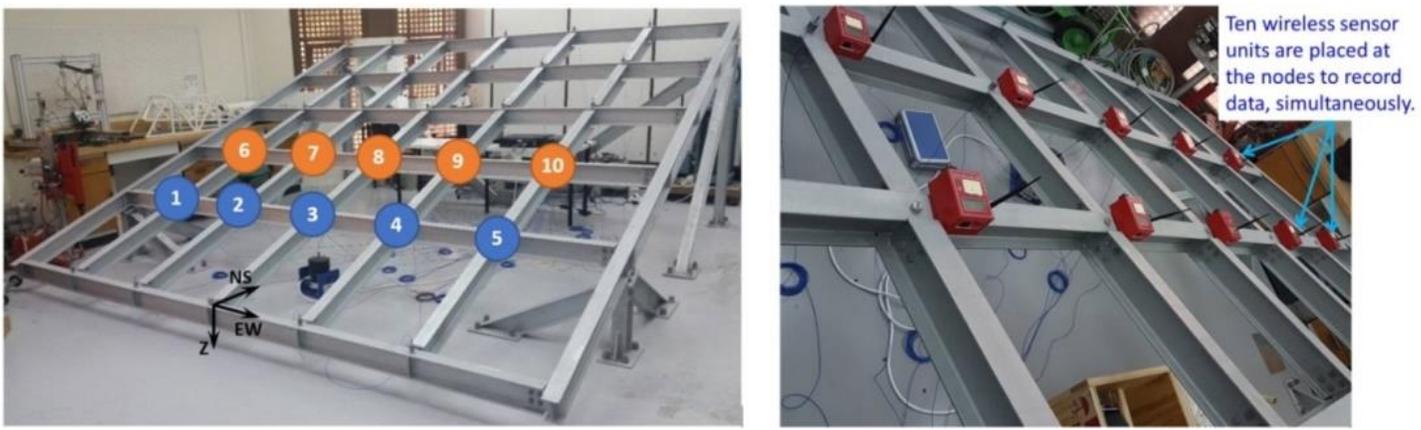

Figure 43 – The test setup and wireless sensors used in Avci et al. [16] and [202]

It was noticed that the process of generating the data required to train the 1D CNNs in [185] and [16] requires a large number of measurement sessions especially in large civil structure. Therefore, Abdeljaber et al. [194] and Avci et al. [195] developed another 1D CNN-based damage quantification approach that requires less effort for training the CNN classifiers. This approach was successfully tested over the data provided under benchmark study of [203]. The adaptive 1D CNN discussed [194] and [195] was processed in C++ installed on a personal computer with Intel ® OpenMP API, used for shared memory multiprocessing. The CNN training and testing were also run on the same PC with 32GB memory. 37 undamaged and 112 damaged frames were utilized to train each of the 12 CNNs. It was noted that the average time for one back-propagation iteration per frame was approximately 150 msec; and the total training time for all CNNs was about 42 sec. A MATLAB code was implemented so that the accelerometer readings for nine structural cases were arranged into frames, normalized, and fed into the CNN classifier to compute the Probability of Damage ($PoD_i$) value at each sensor location. The process is described in Figure 44. The $PoD_i$ computation for a 300 sec long signal with the corresponding CNN classifier was reported to be 5 msec. In other words, the time required to get the $PoD_{avg}$ for 12 acceleration signals was only 60 msec. From another perspective, for each 1 sec of acceleration recording, a processing time of 60/300=0.2 msec was needed to get $PoD_{avg}$ value which indicates the overall health of the structure. Interestingly, this means that the overall SDD speed of this study is about 5000x faster than the "real-time" requirement. Based on the studies [194] and [195], although the CNN classifiers were trained with relatively small number of samples extracted from two damage cases only (out of 10 total cases), they have satisfactorily predicted the structural health corresponding to all damage scenarios.

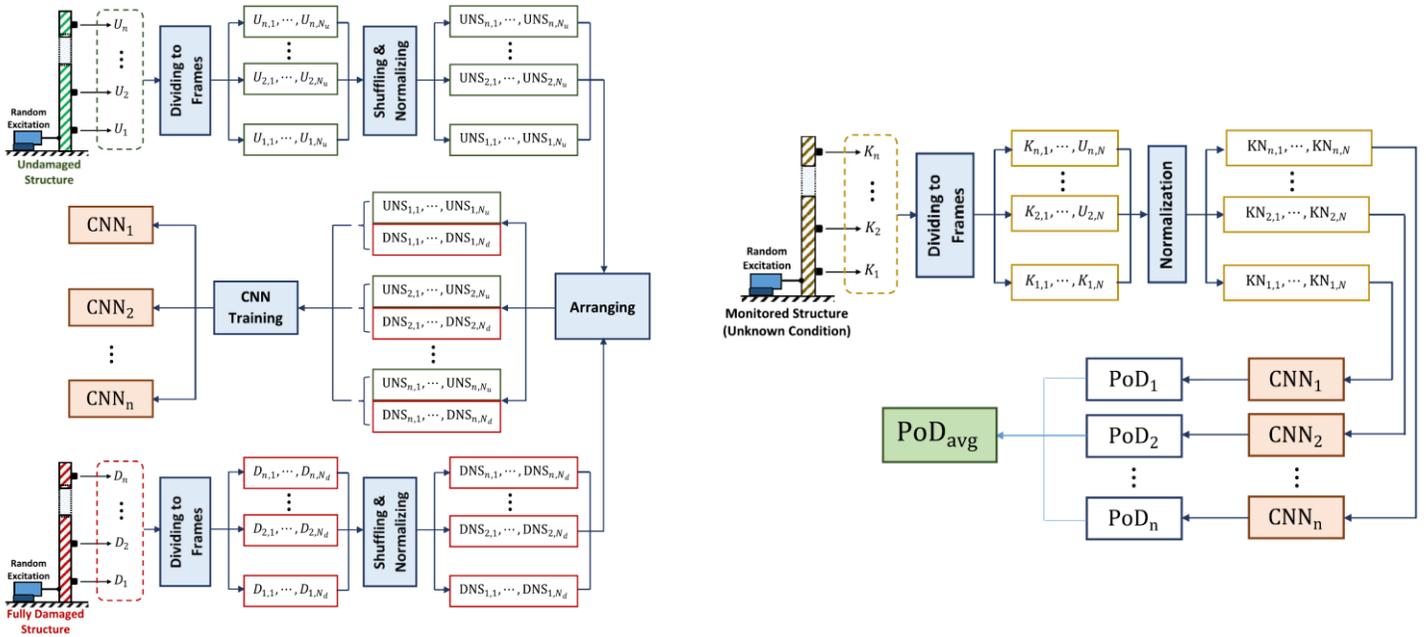

Figure 44 - The data generation and CNN training process (left); $PoD_{avg}$ computation using the trained CNNs (right) in Abdeljaber et al. [194].

These pioneer studies have shown that compact 1D CNNs have proven to be able to accurately distinguish very complex and uncorrelated acceleration recordings. For example in [185], using an ordinary desktop computer, the performance 1D CNNs was tested over a very large laboratory frame (QU grandstand simulator of [204]). The vibration response of the laboratory frame under 31 damage cases was measured and used to train an individual 1D CNN for every particular accelerometer location. Each 1D CNN was only responsible for processing the local data measured at the corresponding location. The performance of this 1D CNN-based damage detection method was tested under a large number of single and double damage cases. The computational complexity analysis was also conducted to estimate the computational time required for the 1D CNNs to process the measured signals. It was reported that the proposed method successfully detected and located the damage for 100% of the damage scenarios. It was indeed the first time that compact 1D CNNs have proven to be able to accurately distinguish uncorrelated and intricate acceleration time-histories which can even challenge a professional in the field with the two damaged and undamaged samples shown in Figure 45. With an ordinary PC, when the performance of the proposed technique was tested, even with a slight stiffness change such as loosened bolts, all the damaged joints were detected without any misses or false alarms with detection speed 45x faster than real-time speed.

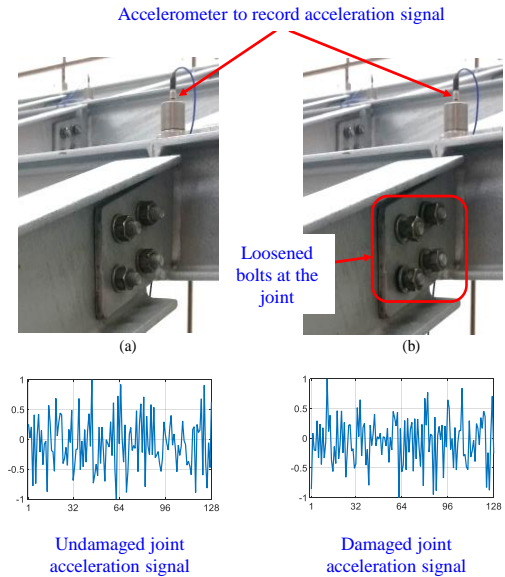

Figure 45 - Damaged and undamaged vibration signals measured at the joints in Abdeljaber et al. [185]

## 6. Conclusions and recommendations for future work

Structural monitoring has been a constant focus of research since the health and safety of structures depends on monitoring the occurrence, formation and propagation of structural damage. Numerous studies have been published on vibration-based structural damage detection methods to identify and quantify structural damage. This paper presented a comprehensive literature review of vibration-based structural damage detection (SDD) methods used for civil engineering structures. This review has covered a broad range of publications on both non-Machine-Learning and Machine-Learning based studies, categorized for parametric and nonparametric applications. Early in the text, a general background on Artificial Intelligence (AI), Machine Learning (ML) and Deep Learning (DL) methods are presented followed by a categorization and review of non-ML based parametric and nonparametric vibration-based SDD methods. Then, the applications of conventional ML-based algorithms in both parametric and nonparametric methods are reviewed. The recent applications of 2D (deep) and 1D (compact) CNNs utilized for vibration-based SDD in civil structures are presented and discussed. The focus is particularly drawn on 1D CNNs due to their state-of-the-art performance levels and numerous computational advantages they can offer.

While the transition from the traditional methods to ML and DL methodologies are presented in this paper, a broad range of non-ML and ML-based parametric and non-parametric approaches are also reviewed in detail. A particular attention was drawn to the techniques applied for extracting damage-sensitive features from the acceleration response of civil structures and also to the ML classifiers used for classifying the extracted features. The reviewed studies were compared in terms of extracted features, classification method, test structure, applied excitation, and type of damage. Based on the literature review conducted in this paper, the following can be summarized as major conclusions:

- A majority of ML-based methods are simply about two tasks: feature extraction and feature classification, which makes the ML-based methods more generic and advantageous than non-ML based methods in vibration-based structural damage detection in civil structures.

- Most studies on ML applications for vibration-based damage detection relied on modal characteristics (natural frequencies, damping ratios, mode shapes) as damage-sensitive features. Yet, recent investigations suggested that such features might not be a good option since they are affected by factors other than structural damage such as changes in temperature and moisture conditions. Another problem associated with modal properties is their low sensitivity to certain types of structural damage. It is, therefore, recommended to avoid using modal properties as damage-sensitive features in ML-based global damage detection approaches.

- Both nonparametric and parametric vibration-based damage detection methods that utilize conventional ML algorithms require extracting a fixed set of hand-crafted features, which are then used as the input by a ML classifier. Numerous feature/classifier combinations have been examined in an attempt to find the best

- combination capable of optimally characterizing structural damage. Nevertheless, it is still ambiguous that a particular combination of features/classifiers will be suitable for any kind of civil structure and for all types of damage scenarios. Moreover, the techniques usually depend on those hand-crafted features such as model identification, Principle Component Analysis (PCA) and Auto-Regressive (AR) modeling, all of which require significant computational time and effort.

- Researchers have recently started to apply nonconventional DL algorithms to develop new vibration-based damage detection techniques that do not require manual feature extraction. Recent studies have shown that both 2D and 1D CNNs are very promising based on their ability to detect and locate damage directly through the raw acceleration time-histories without any need for data preprocessing or hand-crafted feature extraction.

- 1D CNNs are easier to train and have lower computational complexity than their 2D counterparts. In particular, compact 1D CNNs are preferable specifically for cases with limited amount of 1D vibration signals in a dedicated (or isolated) application.

- All damage detection techniques reviewed under ML applications are based on supervised ML algorithms, which require labeled data for training. In other words, the training process requires vibration data collected for the undamaged structure as well as the data measured under several structural damage scenarios. In civil structures, the data-sets for pre-damage and post-damage cases are rarely available. Hence, researchers should come up with innovative solutions for training supervised ML classifiers in the absence of measured damaged data.

- One recommended solution is to research the possibility of training the classifiers using real-life data from the undamaged structure together with simulated data for the damage scenarios. The simulated data can be obtained either numerically using an accurate Finite Element (FE) model or experimentally using a downscaled laboratory model of the monitored structure. This would eliminate the need for having damaged data from an otherwise healthy structure.

- Another solution would be to use semi-supervised and/or unsupervised ML and DL algorithms, which can operate on minimally labeled or unlabeled vibration data. Currently, there are only very few attempts to utilize unsupervised approaches within the scope of structural damage detection [80,108,138,205]. More research effort should be directed toward developing new unsupervised vibration-based damage detection methods.